\documentclass{aa}
\newif\iffigure
\figurefalse
\figuretrue

\usepackage{graphicx,natbib,url,twoopt}
\usepackage[varg]{txfonts}           %% A&A font choice
\usepackage{hyperref}              %% for pdflatex
\usepackage{pdfcomment}              %% for popup acronym meanings
\usepackage{acronym}                 %% for popup acronym meanings
\usepackage{comment}
\usepackage{cases}
\usepackage{empheq}                  %Figs. 1,2 
\usepackage{here}
\usepackage{bm}
\usepackage[version=3]{mhchem}
\hypersetup{% hyperref option
 colorlinks=true,%
 linkcolor=cyan,
 citecolor=cyan,
}

\usepackage{lineno}
\newcommand*\patchAmsMathEnvironmentForLineno[1]{
  \expandafter\let\csname old#1\expandafter\endcsname\csname #1\endcsname
  \expandafter\let\csname oldend#1\expandafter\endcsname\csname end#1\endcsname
  \renewenvironment{#1}
     {\linenomath\csname old#1\endcsname}
     {\csname oldend#1\endcsname\endlinenomath}}
\newcommand*\patchBothAmsMathEnvironmentsForLineno[1]{
  \patchAmsMathEnvironmentForLineno{#1}
  \patchAmsMathEnvironmentForLineno{#1*}}
\AtBeginDocument{
\patchBothAmsMathEnvironmentsForLineno{equation}
\patchBothAmsMathEnvironmentsForLineno{align}
\patchBothAmsMathEnvironmentsForLineno{flalign}
\patchBothAmsMathEnvironmentsForLineno{alignat}
\patchBothAmsMathEnvironmentsForLineno{gather}
\patchBothAmsMathEnvironmentsForLineno{multline}
}
\bibpunct{(}{)}{;}{a}{}{,}    %% natbib cite format used by A&A and ApJ

\makeatletter
\newcommand{\bibnote}[2]{\global\@namedef{#1note}{#2}}
\newcommand{\biblink}[2]{\global\@namedef{#1link}{#2}}
\newcommand{\Tabref}[1]{Table~\ref{#1}}
\newcommand{\Equref}[1]{Eq.~(\ref{#1})}
\newcommand{\Figref}[1]{Fig.~\ref{#1}}
\makeatother

\makeatletter
 \newcommandtwoopt{\citeads}[3][][]{%
   \nonstopmode%              %% fix to not stop at error message in latex
   \href{http://adsabs.harvard.edu/abs/#3}%
        {\def\hyper@linkstart##1##2{}%
         \let\hyper@linkend\@empty\citealp[#1][#2]{#3}}%   %% Rutten, 2000
   \biblink{#3}{\href{http://adsabs.harvard.edu/abs/#3}{ADS}}%
   \errorstopmode}            %% fix to resume stopping at error messages 
 \newcommandtwoopt{\citepads}[3][][]{%
   \nonstopmode%              %% fix to not stop at error message in latex
   \href{http://adsabs.harvard.edu/abs/#3}%
        {\def\hyper@linkstart##1##2{}%
         \let\hyper@linkend\@empty\citep[#1][#2]{#3}}%     %% (Rutten 2000)
   \biblink{#3}{\href{http://adsabs.harvard.edu/abs/#3}{ADS}}%
   \errorstopmode}            %% fix to resume stopping at error messages
 \newcommandtwoopt{\citetads}[3][][]{%
   \nonstopmode%              %% fix to not stop at error message in latex
   \href{http://adsabs.harvard.edu/abs/#3}%
        {\def\hyper@linkstart##1##2{}%
         \let\hyper@linkend\@empty\citet[#1][#2]{#3}}%     %% Rutten (2000)
   \biblink{#3}{\href{http://adsabs.harvard.edu/abs/#3}{ADS}}%
   \errorstopmode}            %% fix to resume stopping at error messages 
 \newcommandtwoopt{\citeyearads}[3][][]{%
   \nonstopmode%              %% fix to not stop at error message in latex
   \href{http://adsabs.harvard.edu/abs/#3}%
        {\def\hyper@linkstart##1##2{}%
         \let\hyper@linkend\@empty\citeyear[#1][#2]{#3}}%  %% 2000
   \biblink{#3}{\href{http://adsabs.harvard.edu/abs/#3}{ADS}}%
   \errorstopmode}            %% fix to resume stopping at error messages 
\makeatother

\newacro{ADS}{Astrophysics Data System}
\newacro{NLTE}{non-local thermodynamic equilibrium}
\newacro{NASA}{National Aeronautics and Space Administration}

\begin{document}
%\linenumbers
\titlerunning{short title}
   \title{Influences of three-dimensional gas flow induced by protoplanets on pebble accretion --$\rm\,I\,$. shear regime}

   \subtitle{}

          \author{Ayumu Kuwahara\inst{1}
          \thanks{\email{kuwahara.a.aa@m.titech.ac.jp}} 
          \and Hiroyuki Kurokawa\inst{2}}

   \institute{Department of Earth and Planetary Sciences, Tokyo Institute of Technology, Ookayama, Meguro-ku, Tokyo, 152-8551, Japan
         \and
             Earth-Life Science Institute, Tokyo Institute of Technology, Ookayama, Meguro-ku, Tokyo, 152-8550, Japan}

   \date{Received September XXX; accepted YYY}

% \abstract{}{}{}{}{} 
% 5 {} token are mandatory
 
  \abstract
 {The pebble accretion model has the potential to explain the formation of various types of planets. The main difference from the planetesimal accretion model is that pebbles not only experience the gravitational interaction with the growing planet, but also gas drag force from the surrounding protoplanetary disk gas. }
   {A growing planet embedded in a disk induces three-dimensional (3D) gas flow, which may influence  pebble accretion. However, so far the conventional pebble accretion model has only been discussed in the unperturbed (sub-)Keplerian shear flow. In this study, we investigate the influence of the 3D planet-induced gas flow on pebble accretion.}
   {Assuming a non-isothermal, inviscid gas disk, we perform 3D hydrodynamical simulations on the spherical polar grid, which has a planet located at its center. Then we numerically integrate the equation of motion of pebbles in 3D using hydrodynamical simulations data. }
   {We find that the trajectories of pebbles in the planet-induced gas flow differ significantly from those in the unperturbed shear flow for a wide range of pebble sizes investigated (${\rm St}=10^{-3}$--$10^{0}$, where ${\rm St}$ is the Stokes number). The horseshoe flow and outflow of the gas alter the motion of the pebbles, which leads to the reduction of the width of the accretion window, $w_{\rm acc}$, and the accretion cross section, $A_{\rm acc}$. On the other hand, the changes in trajectories also cause an increase in relative velocity of pebbles to the planet, which offsets the reduction of $w_{\rm acc}$ and $A_{\rm acc}$. As a consequence, in the Stokes regime, the accretion probability of pebbles, $P_{\rm acc}$, in the planet-induced gas flow is comparable to that in the unperturbed shear flow except when the Stokes number is small, ${\rm St}\sim10^{-3}$, in 2D accretion, or when the thermal mass of the planet is small, $m=0.03$ in 3D accretion. In contrast, in the Epstein regime, $P_{\rm acc}$ in the planet-induced gas flow becomes smaller than that in the shear flow in the Stokes regime in both 2D and 3D accretion, regardless of assumed ${\rm St}$ and $m$.}
   {Our results suggest that the 3D planet-induced gas flow may be helpful to explain the distribution of exoplanets as well as the architecture of the solar system. }


   \maketitle
%

%---------------------------------------------------------
%---------------------------------------------------------
%---------------------------------------------------------
\section{Introduction}\label{sec:Introduction}
More than two decades have passed since the first discovery of an exoplanet \cite[]{Mayor:1995}, and the number of confirmed exoplanets has now reached around 4000, and it continues to increase \footnote{http://exoplanet.eu/catalog/}. As the number of detections of exoplanets increases, the period-mass/radius distribution has shown non-uniformity in the occurrence rate of exoplanets. About half of Sun-like stars harbor close-in super-Earths with orbital periods of less than 85 days, radii of 1--4 $R_{\oplus}$ (Earth radii), and masses of 2--20 $M_{\oplus}$ \cite[Earth masses;][]{Fressin:2013,Weiss:2014}. About 1\% of Sun-like stars host hot Jupiters ($\lesssim$0.1 au), while around 10\% stars harbor cold Jupiters ($\sim$1--10 au) with a peak of occurrence rate between 2--3 au \cite[]{Johnson:2010,Fernandes:2019}. The occurrence rate of the planets with an orbital distance of $\sim$30--300 au are as low as 1\% at most \cite[]{Bowler:2016}. Discussions about the origin of exoplanetary systems are still ongoing.

Planets are formed in protoplanetary disks. One of the major planet formation theories that are currently being considered is the planetesimal accretion model. In this model, the building blocks of the planets are km-sized planetesimals.  As the planetary embryos accrete planetesimals and become massive enough, their gravitational focusing becomes more significant, and then the massive embryos grow faster than other smaller ones. This phase is called the runaway growth process \cite[]{Kokubo:1996}. After the runaway growth phase, the largest embryos (protoplanets) grow in an oligarchic manner, while most planetesimals remain small: this stage is called the oligarchic growth process \cite[]{Kokubo:1998}. The disk gas accretes onto a core formed in this way, and the total mass of the planet exceeds critical core mass ($\sim10$ $M_{\oplus}$), runaway gas accretion is triggered and the planet evolves into a gas giant \cite[e.g.,][]{Mizuno:1980,Pollack:1996,Ikoma:2000}. In the planetesimal accretion model, however, it is difficult to form gas giants within the lifetime of the disk at distances larger than $\sim5$ au, unless the disk was massive \cite[e.g.,][]{Kobayashi:2011}.

Pebble accretion is a new model of planet formation \cite[e.g.,][]{Ormel:2010,Ormel:2012,Lambrechts:2012,Lambrechts-Johansen:2014,Lambrechts:2014,Guillot:2014,Ida:2016}, which may overcome several problems remaining in the theory based on planetesimal accretion \cite[e.g.,][]{Kokubo:2000}. Accreting mm--cm-sized particles (pebbles) onto the proto-cores can form cores massive enough to trigger gas giant formation within the lifetime of gas disks \cite[]{Lambrechts:2012,Levison:2015,Tanaka:2019,Bitsch:2019}. In addition to the formation of gas giants' cores, the pebble accretion scenario has been applied to the size distribution of terrestrial planets in the solar system \cite[]{Morbidelli:2015,Drkazkowska:2016}, water delivery to the terrestrial planets \cite[]{Morbidelli:2016,Sato:2016,Ida:2019}, the preference for prograde spin of the major planetary bodies in the solar system \cite[]{Johansen:2010,Visser:2019}, the formation of close-in super-Earths \cite[]{Chatterjee:2014,Chatterjee:2015,Moriarty:2015,Lambrechts:2019,Izidoro:2019,Bitsch:2019}, and the mass distribution of the planets around cool dwarf stars \cite[]{Ormel:2017,Schoonenberg:2019}. Pebble accretion in inviscid disks may explain the dichotomy between the inner super-Earths and outer gas giants \cite[]{Fung:2018}. However, the final configurations of the planetary systems strongly depends on the pebble flux \cite[]{Bitsch:2019}. Even with the pebble accretion model, the origin of the non-uniformity of the distribution of exoplanets remains obscure as if it was enveloped in fog.

A key parameter to control the outcome of the planet formation models is the accretion probability of pebbles or smaller dust onto a growing body, which has been evaluated mostly in the Keplerian or sub-Keplerian shear flow around a protoplanet \cite[]{Ormel:2010,Lambrechts:2012,Lambrechts-Johansen:2014,Sellentin:2013,Guillot:2014,Ida:2016,Visser:2016}. The loss of angular momentum due to the gas drag force leads to efficient accretion of pebble-sized particles compared to planetesimals. The particle scale height, which depends on the turbulence and particle size, has also been found to be important as it determines whether pebbles accrete three- or two-dimensionally.

Recent studies have reported the detailed 3D structure of the gas flow induced by a planet embedded in a protoplanetary disk \cite[]{Ormel:2015b,Fung:2015,Lambrechts:2017,Cimerman:2017,Kurokawa:2018,Kuwahara:2019,Bethune:2019}. The horseshoe flows extending of the anterior-posterior direction in the planet's orbital direction have a characteristic vertical structure like a column. A substantial amount of gas from the disk enters the gravitational sphere of the planet (Bondi or Hill sphere) at high latitudes (inflow), and exits through the midplane region of the disk (outflow). 

The induced gas flow may alter the accretion probability of pebbles. \cite{Kuwahara:2019} showed that the speed of midplane outflow increases with the protoplanet mass and eventually exceeds the terminal speed of incoming pebbles. This suggests that the outflow may inhibit accretion, whereas the polar inflow may act as the accretion window when the accretion is 3D. \cite{Popovas:2018a,Popovas:2018b} incorporated dust particles in their hydrodynamical simulations and found that small particles (10 $\mu$m--1 cm) move away from the planet in the horseshoe flow and avoid accreting onto Earth- and Mars-sized planets. \cite{Ormel:2013} also reported the reduction of accreting small particles in 2D flow for an approximately Mars-sized core. From the analytical argument, \cite{Rosenthal:2018a} and \cite{Rosenthal:2018b} proposed that the gas flow around the bound atmosphere may set the ``flow isolation mass''.

Though these studies suggest the importance of considering the gas flow induced by a protoplanet, a comprehensive study for ranges of core masses and pebble sizes is missing. In this study we perform hydrodynamical simulations of the gas flow around an embedded planet and compute the pebble trajectories and their accretion probabilities.

The structure of this paper is as follows. In Sect. \ref{sec:methods} we describe the numerical method. In Sect. \ref{sec:result} we show the results obtained from a series of simulations and comparison between the analytical estimations. In Sect. \ref{sec:discussion} we discuss the implications for planet formation. We summarize in Sect. \ref{sec:conclusion}.
%---------------------------------------------------------
%---------------------------------------------------------
%---------------------------------------------------------
\section{Methods}\label{sec:methods}
 \begin{figure}[htbp]
 \resizebox{\hsize}{!}
 {\includegraphics{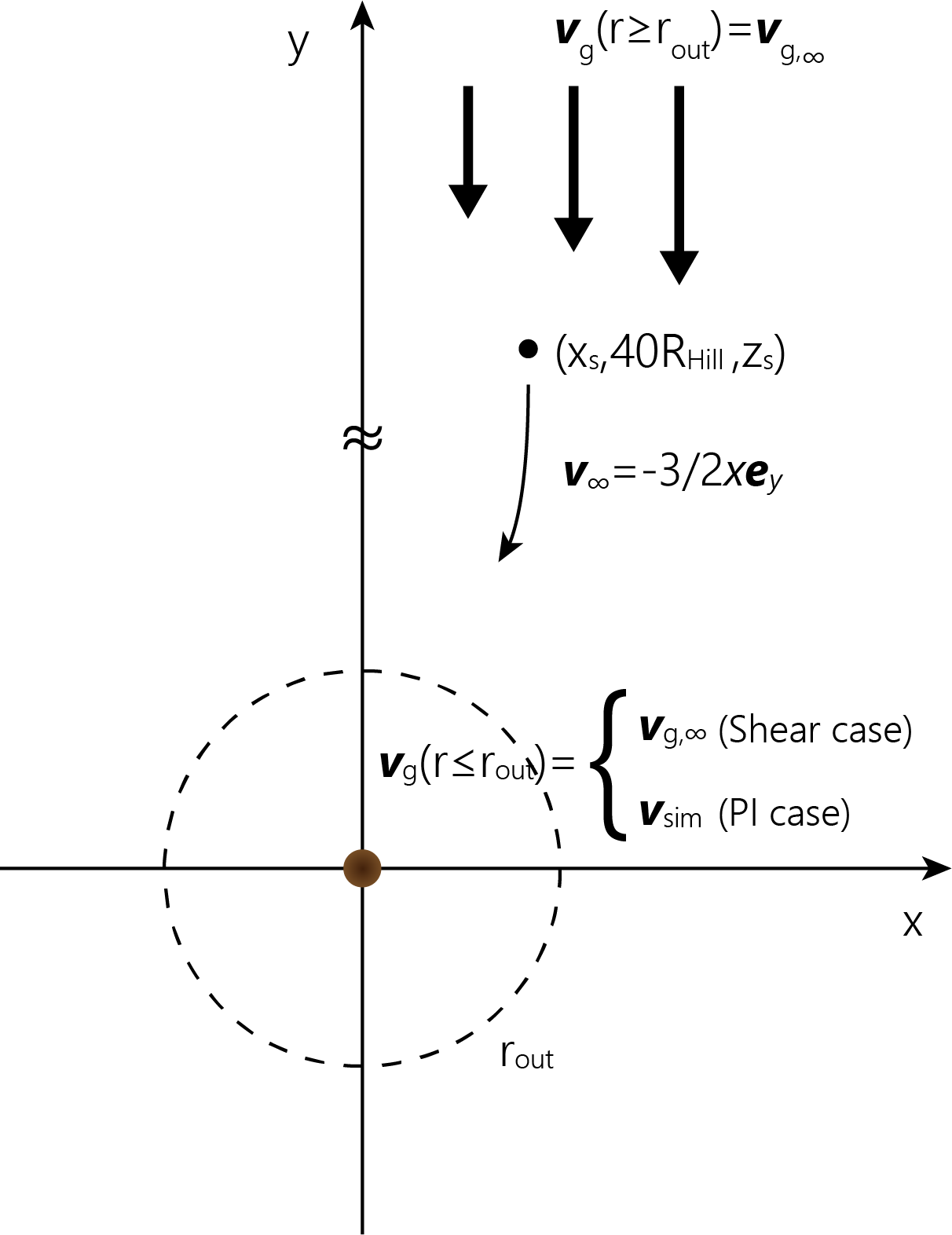}} 
 \caption{Schematic picture of orbital calculation of pebbles. A planet is located at the origin of the co-rotational frame. The dashed line represents the outer boundary of the hydrodynamical simulations. The starting point of orbital calculation is beyond the outer boundary of hydrodynamical simulations. Its $y$-component is fixed at $40R_{\rm Hill}$. The initial velocity of the pebble is the same with the Keplerian shear velocity. The gas velocity is assumed to be the speed of the Keplerian shear both inside and outside of $r_{\rm out}$ in shear flow case in the Stokes regime (\textit{Shear case}), but is switched to the gas velocity obtained from the hydrodynamical simulations within $r_{\rm out}$ in tne planet-induced flow case in the Epstein regime (\textit{PI-Epstein case}), and in the Stokes regime (\textit{PI-Stokes case}).}
\label{fig:schematic-pic}
\end{figure}

\begin{table}[!tp]
\caption{List of hydrodynamical simulations. The following columns give the simulation name, the size of the Bondi radius of the planet, the size of the Hill radius of the planet, the size of the inner boundary, the size of the outer boundary, the length of the calculation time, and the dimensionless thermal relaxation timescale $\beta$, respectively.}
\centering
\begin{tabular}{ccccccc}\hline\hline
Name & $R_{\rm Bondi}$ & $R_{\rm Hill}$ & $r_{\rm inn}$ & $r_{\rm out}$ & $t_{\rm end}$ & $\beta$\\ \hline
\texttt{m003} & 0.03 & 0.22 & 9.32$\times10^{-3}$ & 0.5 & 50 & $0.09$ \\
\texttt{m01} & 0.1 & 0.32 & 1.39$\times10^{-2}$ & 5  & 150 & 1 \\
\texttt{m03} & 0.3 & 0.46 & 2$\times10^{-2}$ & 15 & 200 & 9\\\hline
\end{tabular}
\label{tab:1}
\end{table}

\subsection{Model overview}
Gas flow around an embedded planet is perturbed by the gravity of the planet, and then a 3D flow structure is formed around the planet. In this study, we call the above-mentioned perturbed gas flow the \textit{planet-induced (PI) gas flow} in contrast to the unperturbed shear flow. To investigate the influence of planet-induced gas flow on pebble accretion, we performed 3D hydrodynamical simulations (Sect. \ref{sec:hydrosimulation}) and then calculated the trajectories of pebbles in the gas flow (Sect. \ref{sec:pebblesimulation}). Finally, we compute the accretion probability of pebbles (Sect. \ref{sec:2.4}). Through all of our simulations, the lengths, times, velocities, and densities are normalized by the disk scale height $H$, the reciprocal of the orbital frequency $\Omega^{-1}$, the sound speed $c_{\rm s}$, and the unperturbed gas density at the location of the planet $\rho_{\rm disk}$, respectively. In this dimensionless unit system, the dimensionless mass of the planet (called ``thermal mass''; \cite{Fung:2015}) is expressed by the ratio of the Bondi radius of the planet, $R_{\rm Bondi}$, to the disk scale height,
\begin{align}
m=\frac{R_{\rm Bondi}}{H}=\frac{GM_{\rm pl}}{c_{\rm s}^{3}\Omega},
\end{align}
where $G$ is the gravitational constant, and $M_{\rm pl}$ is the mass of the planet. The Hill radius is given by $R_{\rm Hill}=(m/3)^{1/3}$ in this unit. When we assume a solar-mass host star and a disk temperature profile $T=270\ (a/1\ {\rm au})^{-1/2}$ K, which corresponds to the minimum mass solar nebula (MMSN) model \cite[]{Weidenschilling:1977,Hayashi:1985}, $M_{\rm pl}$ is described by 
\begin{align}
M_{\rm pl}\simeq12m\left(\frac{a}{1\ {\rm au}}\right)^{3/4}M_{\oplus},\label{eq:planetarymass}
\end{align}
where $a$ is the orbital radius \cite[]{Kurokawa:2018}.

\subsection{3D hydrodynamical simulations}\label{sec:hydrosimulation}
In this study, we performed non-isothermal 3D hydrodynamical simulations of the gas of the protoplanetary disk around a planet. Our simulations were performed in a spherical polar coordinate co-rotating with a planet with Athena++ \citep[][Stone et al. in prep.]{White:2016}. Most of our methods of hydrodynamical simulations are the same as described in detail in \cite{Kurokawa:2018}, but include several differences. Here we summarize the differences with \cite{Kurokawa:2018} as below. 

We ignored the headwind of the gas for all of our simulations. The external force in the Euler equation does not contain the global pressure force due to the sub-Keplerian motion of the gas. This assumption is justified because we focus on the shear regime of pebble accretion. There are two regimes of pebble accretion: the headwind regime and shear regime \cite[]{Lambrechts:2012,Ormel:2017}\footnote{In this study, we used ``headwind'' and ``shear'' regimes as the names to distinguish the pebble accretion regimes, which are used in \cite{Ormel:2017}. These regimes are referred to as ``Bondi'' and ``Hill'' regimes in \cite{Lambrechts:2012}.}. In the headwind regime, the approach speed of the pebble to the planet is dominated by the headwind of the gas,
\begin{align}
v_{\rm hw}=\eta v_{\rm K},\label{eq:vhw}
\end{align}
where 
\begin{align}
\eta = -\frac{1}{2}\left(\frac{c_{\rm s}}{v_{\rm K}}\right)^{2}\frac{\mathrm{d}\ln P}{\mathrm{d}\ln a}\label{eq:eta}
\end{align}
is a dimensionless quantity characterizing the pressure gradient of the disk gas and $v_{\rm K} = a\Omega$ is the Kepler velocity, and $P$ is the pressure of the gas. On the other hand, the approach speed is dominated by the Keplerian shear in the latter case. 
The transition from the headwind to the shear regime occurs when the mass of the planet exceeds the transition mass given by 
\begin{align}
M_{\rm t} =\sqrt{\frac{1}{3}}\frac{v_{\rm hw}^{3}}{G\Omega}\simeq2.0\times10^{-4}M_{\oplus}\left(\frac{a}{1\ {\rm au}}\right)^{3/2}\left(\frac{v_{\rm hw}}{30\ {\rm m\ s^{-1}}}\right)^{3},\label{eq:transition}
\end{align}
\cite[]{Lambrechts:2012,Johansen:2017}. From \Equref{eq:planetarymass}, \Equref{eq:transition} can be rewritten as 
\begin{align}
m\simeq1.7\times10^{-5}\,\left(\frac{a}{\rm 1\, au}\right)^{3/4}\left(\frac{v_{\rm hw}}{30\ {\rm m\ s^{-1}}}\right)^{3}.\label{eq:dim-less-transition}
\end{align}
Since the assumed planetary masses in this study are larger than the transition mass (\Equref{eq:dim-less-transition}), we focus on the shear regime. We note that the effect of headwind on pebble drift is considered when we compute the accretion probability of pebbles (Sect. \ref{sec:2.4}).

\cite{Kurokawa:2018} fixed the size of the inner boundary for all of their simulations, but we varied it according to the mass of the planet. Assuming the density of the embedded planet $\rho_{\rm pl}=5$ g/cm$^{3}$ leads to the physical radius of the planet, $R_{\rm pl}$, as given by
\begin{align}
R_{\rm pl}&=\left(\frac{3M_{\rm pl}}{4\pi\rho_{\rm pl}}\right)^{1/3}\nonumber \\
&=\left(\frac{9M_{\ast}}{4\pi\rho_{\rm pl}}\right)\frac{R_{\rm Hill}}{a}\ \ \left(\because\ R_{\rm Hill}=a\left(\frac{M_{\rm pl}}{3M_{\ast}}\right)^{1/3}\right)\nonumber \\
&\simeq3\times10^{-3}m^{1/3}\ \Biggl(\frac{\rho_{\rm pl}}{5\ {\rm g/}{\rm cm}^{3}}\Biggr)^{-1/3}\ \Biggl(\frac{M_{\ast}}{1\ M_{\odot}}\Biggr)^{1/3}\ \Biggl(\frac{a}{1\ {\rm au}}\Biggr)^{-1},\label{eq:planetaryradius}
\end{align}
where $M_{\ast}$ and $M_{\odot}$ are the stellar mass and the solar mass. Our hydrodynamical simulations are computationally expensive, but the CPU time is reduced when we increase the size of the inner boundary. Therefore, we regard the size of $r_{\rm inn}$ as being determined by \Equref{eq:planetaryradius} with $a=0.1$ au, though we consider pebble accretion not only at 0.1 au, but also at various orbital radii. We confirmed that the size of the inner boundary does not affect our results.

Our simulations utilized the $\beta$ cooling model, where the temperature $T$ relaxes toward the background temperature $T_{0}$ with the dimensionless timescale $\beta$ \cite[e.g.,][]{Gammie:2001}. To determine an adequate $\beta$ value for a certain planetary mass, we followed the discussion in \cite{Kurokawa:2018}. Considering the relaxation time $\beta$ for the temperature perturbation whose wavelength is equal to the Bondi radius of the planet, we fixed $\beta=1$ at $m=0.1$ \cite[]{Malygin:2017} and assumed that $\beta$ scales linearly with the square of $m$, namely, $\beta=\left(m/0.1\right)^{2}$ (see \cite[]{Kurokawa:2018} for the discussion).

We listed our parameter sets in \Tabref{tab:1}. The range of the dimensionless planetary masses, $m=0.03$--0.3, corresponds to a three Mars-masses to super-Earth-sized planet, $M_{\rm pl}=0.36$--3.6 $M_{\oplus}$, orbiting a solar-mass star at 1 au. Since the rapid increase of the gravity of the planet in the unperturbed disk affects the results of the simulations \cite[]{Ormel:2015a}, the gravity of the planet is gradually inserted into the disk at the injection time, $t_{\rm inj}=0.5$. The gravity term is given by \cite[]{Ormel:2015a,Ormel:2015b,Kurokawa:2018,Kuwahara:2019}
\begin{align}
\bm{F}_{\rm p}=-\nabla\left(\frac{m}{\sqrt{r^{2}+r_{\rm s}^{2}}}\right)\left\{1-\exp\left[-\frac{1}{2}\left(\frac{t}{t_{\rm inj}}\right)^{2}\right]\right\},
\end{align}
where {$r$ is the distance from the center of the planet and \ $r_{\rm s}$ is the smoothing length. We assumed $r_{\rm s}=0.1m$ \cite[]{Kurokawa:2018}. The numerical resolution is $[\log r,\ \theta,\ \phi]=[128, 64, 128]$. We adopted a logarithmic grid for the radial coordinates, which has a higher resolution in the vicinity of the planet.

\subsection{3D Orbital calculation of pebbles}\label{sec:pebblesimulation}
We calculated the trajectories of pebbles influenced by the planet-induced gas flow in the frame co-rotating with the planet (\Figref{fig:schematic-pic}). We performed three types of orbital calculations: one is performed in the unperturbed shear flow (the gas has an initial condition profile), and the others are performed with the hydro-simulations data. We explain the detail of the orbital calculations in Sect. \ref{sec:orbital-setting}. The origin of the system is located at the position of the embedded planet. In order to investigate the effect of planet-induced gas flow on the motion of pebbles, we performed parameter studies for three planet masses, $m=0.03, 0.1$, and 0.3, and various pebble sizes using the numerical results obtained from the above-mentioned hydrodynamical simulations. We used the final state of the hydro-simulations data ($t=t_{\rm end}$), where the flow field seems to have reached the steady state.
\subsubsection{Equation of motion}
The dimensionless equation of motion of a pebble is described by
\begin{align}
\frac{\mathrm{d}\bm{v}}{\mathrm{d}t}=
\begin{pmatrix}
2v_{y}+3x \\
-2v_{x} \\
0
\end{pmatrix}
-\frac{m}{r^{3}}
\begin{pmatrix}
x \\
y\\
z
\end{pmatrix}
+\bm{F}_{\textrm{drag}},\label{eq:EOM}
\end{align}
where $\bm{v}=(v_{x},v_{y},v_{z})$ is velocity of the pebble and $(x,y,z)$ is the coordinates of the pebble \cite[]{Ormel:2010,Visser:2016}. The first and second term on the  right-hand side of \Equref{eq:EOM} are the Coriolis and tidal forces and the two-body interaction force with the embedded planet. The third term is the gas drag force acting on the pebble expressed by
\begin{align}
\bm{F}_{\rm drag} = -\frac{\bm{v}-\bm{v}_{\rm g}}{\rm St},\label{eq:gasdrag}
\end{align}
where $\bm{v}_{\rm g}$ is the gas velocity, and St is the dimensionless stopping time of a pebble, called the Stokes number, ${\rm St}=t_{\rm stop}\Omega$. We assumed ${\rm St}=10^{-3}$--1.

We omitted the dimensionless vertical component of the tidal force, $-z\bm{e}_{z}$ in \Equref{eq:EOM}. Through the test integrations with this term, we found that most pebbles settled in the midplane region of the disk before they enter the calculation domain of hydrodynamical simulations when ${\rm St}\gtrsim0.1$. In reality, turbulent stirring would balance the vertical component of the tide on statistical average. The sources of this turbulence will be discussed in Sect. \ref{sec:turbulence}. Though our hydrodynamical simulations do not resolve the turbulence, we assumed that the turbulent and tidal forces in the vertical direction canceled each other out. Considering the effect of random motion due to the turbulence \cite[]{Xu:2017,Picogna:2018} is beyond the scope of this study.

The gas drag force is divided into two regimes: the Epstein and the Stokes regimes depending on the relationship between the size of the particle and the mean free path of the gas. The stopping time of the particle in each regime is described by
\begin{empheq}[left={t_{\rm stop}=\empheqlbrace}]{alignat=2}
&\displaystyle\frac{\rho_{\bullet}s}{\rho_{\rm g}c_{\rm s}}, \quad &(\text{Epstein regime}: s\leq\frac{9}{4}\lambda)\label{eq:Epstein-regime}\\
&\displaystyle\frac{4\rho_{\bullet}s^{2}}{9\rho_{\rm g}c_{\rm s}\lambda},  \quad &(\text{Stokes regime}: s\geq\frac{9}{4}\lambda)\label{eq:Stokes-regime}
\end{empheq}
where $\rho_{\bullet}$ is the internal density of the pebble, $s$ is the radius of the pebble, $\rho_{\rm g}$ is the density of the gas, and $\lambda$ is the mean free path of the gas, $\lambda=\mu m_{\rm H}/\rho_{\rm g}\sigma_{\rm mol}=1.44\ {\rm cm}\ (a/{\rm 1\ au})^{11/4}$ with $\mu$, $m_{\rm H}$, and $\sigma_{\rm mol}$ being the mean molecular weight, $\mu=2.34$, the mass of the proton, and the molecular collision cross section, $\sigma_{\rm mol}=2\times10^{-15}\,{\rm cm}^{2}$ \cite[]{Chapman:1970,Weidenschilling:1977a,Nakagawa:1986}. The gas density at the midplane is given by $\rho_{\rm g}=\Sigma_{\rm g}/\sqrt{2\pi}H$, where $\Sigma_{\rm g}$ is the gas surface density, $\Sigma_{\rm g}=1700\,{\rm g\,cm}^{-2}\,(a/{\rm 1\,au})^{-3/2}$. The pebble size in the MMSN can be expressed by 
\footnotesize
\begin{empheq}[left={s=\empheqlbrace}]{alignat=2}
&\displaystyle3.4\times10^{2}\,{\rm cm}\ {\rm St}\,\Biggl(\frac{\rho_{\bullet}}{2\,{\rm g\ cm}^{-3}}\Biggr)^{-1}\Biggl(\frac{a}{1\,{\rm au}}\Biggr)^{-3/2},\label{eq:pebble-size1}\quad(\text{Epstein regime})\\
&\displaystyle33.2\,{\rm cm}\,\sqrt{{\rm St}}\,\Biggl(\frac{\rho_{\bullet}}{2\,{\rm g\,cm}^{-3}}\Biggr)^{-1/2}\Biggl(\frac{a}{1\,{\rm au}}\Biggr)^{5/8},\label{eq:pebble-size2}\quad\quad(\text{Stokes regime})
\end{empheq}
\normalsize 
\cite[]{Lambrechts:2012}. From Eqs. (\ref{eq:Epstein-regime})--(\ref{eq:pebble-size2}), the transition from the Stokes to the Epstein regime occurs when
\begin{align}
{\rm St}=9.5\times10^{-3}\,\Biggl(\frac{\rho_{\bullet}}{2\,{\rm g\ cm}^{-3}}\Biggr)\,\Biggl(\frac{a}{\rm 1\,au}\Biggr)^{17/4}.\label{eq:pebble-transition}
\end{align}

Since the mean free path of the gas is proportional to the reciprocal of the gas density, $\lambda\propto\rho_{\rm g}^{-1}$, the stopping time in the Stokes regime is independent of the gas density. In our model, the Stokes number, ${\rm St}$, in \Equref{eq:EOM} is independent of the local gas density, $\rho_{\rm g}$, in the Stokes regime, whereas ${\rm St}$ is proportional to $\rho^{-1}_{\rm g}$ in the Epstein regime. The gas density increases significantly in the vicinity of the planet, then the effective Stokes number decreases as the gas density increases. The threshold value of St for the transition varies as a function of the distance from the star (\Equref{eq:pebble-transition}). Because we discuss planet formation at various orbital radii, both the Epstein and Stokes regimes are considered for a range of ${\rm St}=10^{-3}$--$10^{0}$.

\subsubsection{Numerical settings: 3D orbital calculation}\label{sec:orbital-setting}
We integrated \Equref{eq:EOM} using a fifth-order Runge-Kutta-Fehlberg method \cite[]{Fehlberg:1969,Shampine:1976,Eshagh:2005}. In this scheme each time step is controlled automatically depending on the relative acceleration of the pebble to the planet. The maximum relative error tolerance was set to $10^{-8}$, which ensures numerical convergence \cite[]{Ormel:2010,Visser:2016}. We fixed the $y$-coordinate of the starting point of pebbles at $|y_{\rm s}|=40R_{\rm Hill}$ \cite[]{Ida:1989}. The $x$- and $z$-coordinates of the starting point of pebbles, $x_{\rm s}$ and $z_{\rm s}$, are the parameters. We assumed that the initial velocity of the pebble is $\bm{v}_{{\rm p},\infty}=(0,-3/2x,0)$. We set the lower limits of $z_{\rm s}$ as $z_{\rm s}=0$ and $x_{\rm s}$ as $x_{\rm s}=0.01b_{x}$ as values sufficiently small to be the common criteria for any planetary mass and Stokes number, which have an order of $\sim10^{-4}$ [H], where $b_{x}$ is the maximum impact parameter of accreted pebbles for the Shear case (Appendix \ref{sec:pebbleaccretion}). The upper limits of $x_{\rm s}$ and $z_{\rm s}$ are set to be sufficiently large values, which have an order of a few disk scale heights.  We integrated \Equref{eq:EOM} for pebbles with their initial spatial intervals of $0.01b_{x}$ in $x$ and $z$ direction. Since the starting point is beyond the calculation domain of our hydrodynamical simulations, we assumed that the gas velocity was set to be Keplerian shear, $\bm{v}_{\rm g}=(0,-3/2x,0)$, in the outer region of the domain, $r\geq r_{\rm out}$. 

We performed three types of orbital calculations: shear flow case in the Stokes regime (\Equref{eq:Stokes-regime}; hereafter \textit{Shear case}), where we adopted unperturbed Keplerian shear flow where the gas density is uniform; planet-induced flow case in the Epstein regime (\Equref{eq:Epstein-regime}; hereafter \textit{PI-Epstein case}); and that in the Stokes regime (hereafter \textit{PI-Stokes case}). The gas density increases significantly in the vicinity of the planet in the PI-cases. Since the Stokes number in the Epstein regime depends on the gas density, the effective Stokes number decreases as the gas density increases. For the latter two cases, we switched the gas flow from Keplerian to planet-induced gas flow obtained by hydrodynamical simulations at $r=r_{\rm out}$.  We interpolated the gas velocity using the bilinear interpolation method (see Appendix \ref{sec:interpolation}).

In the case of the \texttt{m003} run, we found that the horseshoe flow formed unexpected vortices, which influences the pebble trajectories. The origin of these vortices is unknown, but it is likely to be a numerical artifact due to the spherical polar coordinates centered at the planet, in which the resolution becomes too low to resolve the horseshoe flow far from the planet when the assumed planet mass is small. Only in this case we use the limited part of the calculation domain, $r\leq0.3$, to avoid the effects of the vortices. We discuss the width of horseshoe flow in Sect. \ref{sec:HS}.

We assumed that the pebble accreted onto the planet when the distance between the pebble and the planet became smaller than the following critical values: pebbles  accreted when $r\leq 2r_{\rm inn}$ for the Stokes regime, or $r\leq0.3R_{\rm Bondi}$ for the Epstein regime. In the Epstein regime, pebbles sometimes stagnate in the dense region near the planet. This is because the radial velocity of the pebble is so small but takes a non-zero value in the isolated envelope. To finish the calculation we assumed that pebbles accreted onto the planet when they entered the bound atmosphere.

\subsection{Calculation of accretion probability of pebbles}\label{sec:2.4}
\subsubsection{Width of accretion window and accretion cross section}
We defined the width of the accretion window by
\begin{align}
w_{\rm acc}(z)=x_{\rm max}(z)-x_{\rm min}(z),\label{eq:acc-width}
\end{align}  
where $x_{\rm max}(z)$ and $x_{\rm min}(z)$ are the maximum and the minimum value of the $x$-component of the starting point of accreted pebbles at a certain height. In the unperturbed shear flow, the width of the accretion window, $w_{\rm acc}(z)$, is identical to the maximum impact parameter of accreted pebbles, $b_{x}$, when ${\rm St}<1$ as $x_{\rm min}(z)=0$ (\Equref{eq:b-sh} in Appendix \ref{sec:pebbleaccretion}) Using this definition, we defined the accretion cross section as,
\begin{align}
A_{\rm acc}=2\int_{z_{\rm min}}^{z_{\rm max}}\int_{x_{\rm min}(z)}^{x_{\rm max}(z)}dxdz,\label{eq:acc-sec}
\end{align}
where $z_{\rm max}$ and $z_{\rm min}$ are the maximum and the minimum value of the $z$-component of the starting point of accreted pebbles. The factor of 2 in \Equref{eq:acc-sec} comes from the symmetrical structure of the planet-induced gas flow (see Sect. \ref{sec:result1} and \Figref{fig:streamline}). We reduced the spatial intervals stepwise near the edge of the accretion window, and determined $x_{\rm max}(z)$ and $x_{\rm min}(z)$ with sufficient accuracy. We confirmed that obtained $w_{\rm acc}$ has at least 3 significant digits. 

\subsubsection{Accretion probability}\label{sec:accretionprobability}
Here we consider the accretion efficiency of pebbles in terms of the accretion probability of pebbles. We define the accretion probability of pebbles as
\begin{align}
P_{\rm acc}=\frac{\dot{M}_{\rm p}}{\dot{M}_{\rm disk}},\label{eq:Pacc}
\end{align}
where $\dot{M}_{\rm p}$ is the accretion rate of pebbles onto a protoplanet and $\dot{M}_{\rm disk}$ is the radial inward mass flux of pebbles in the gas disk described by 
\begin{align}
\dot{M}_{\rm disk}=2\pi a\Sigma_{\rm p}|v_{\rm drift}|,\label{eq:M-dot-disk}
\end{align}
where $\Sigma_{\rm p}$ is the surface density of pebbles and $v_{\rm drift}$ is the radial drift velocity of pebbles given by \cite{Weidenschilling:1977} as 
\begin{align}
v_{\rm drift}=-\frac{2{\rm St}}{1+{\rm St}^{2}}v_{\rm hw}.\label{eq:vdrift}
\end{align}
The density distribution of pebbles is described by
\begin{align}
\rho_{\rm p}(z)=\frac{\Sigma_{\rm p}}{\sqrt{2\pi}H_{\rm p}}\exp\left[-\frac{1}{2}\left(\frac{z}{H_{\rm p}}\right)^{2}\right],\label{eq:pebble-density}
\end{align}
where $H_{\rm p}$ is the scale height of pebbles \cite[]{Durbrulle:1995,Cuzzi:1993,Youdin:2007}:
\begin{align}
H_{\rm p}=\left(1+\frac{{\rm St}}{\alpha}\frac{1+2{\rm St}}{1+{\rm St}}\right)^{-1/2},\label{eq:pebble-scaleheight}
\end{align}
where $\alpha$ is the turbulent parameter in the disk introduced by \cite{Shakura:1973}. The sources of the turbulence are discussed in Sect. \ref{sec:turbulence}. Our calculation of accretion probability assumed that pebbles have a vertical distribution given by \Equref{eq:pebble-density}. This approach neglects the effect of random motion of individual particles, which is discussed in Sect. \ref{sec:turb}. The accretion rate of pebbles, $\dot{M}_{\rm p}$, divided into two formulas:
\begin{align}
\dot{M}_{\rm p, 2D}&=2\int_{x_{\rm min}(0)}^{x_{\rm max}(0)}\Sigma_{\rm p}\bm{v}_{{\rm p},\infty}dx,\label{eq:M-dot-2D-sim}
\end{align}
in the 2D case, and
\begin{align}
\dot{M}_{\rm p, 3D}&=4\int_{z_{\rm min}}^{z_{\rm max}}\int_{x_{\rm min}(z)}^{x_{\rm max}(z)}\rho_{{\rm p},\infty}(z)\bm{v}_{{\rm p},\infty}dxdz,\label{eq:M-dot-sim}
\end{align}
in the 3D case. In order to account for the accretion from both $x>0$ and from $x<0$, we multiply \Equref{eq:M-dot-2D-sim} by 2 and \Equref{eq:M-dot-sim} by 4, respectively. The accretion probabilities in both 2D and 3D do not depend on the orbital radius, $a$ (see Appendix \ref{sec:appendix-c}).

%Since our disk model is the minimum-mass solar nebula model \cite[]{Weidenschilling:1977,Hayashi:1985}, the aspect ratio of the disk and the typical value of the headwind are expressed by $H/r\simeq0.033\left(a/1\ {\rm au}\right)^{1/4}$, $v_{\rm hw}/c_{\rm s}\simeq0.05\left(a/1\ {\rm au}\right)^{1/4}$. 
The existence of massive pebble reservoirs ($\sim10^{2}\,M_{\oplus}$) in the outer regions of the young disks has been suggested by (sub)millimeter dust observations \cite[]{Ricci:2010,Andrews:2013,Ansdell:2016}.  The dust mass is correlated with the accretion rate onto young stars. Given a typical T Tauri star with a disk accretion rate $10^{-8}\,M_{\odot}$/Myr, where $M_{\odot}$ is the solar mass, the dust mass is $\sim10^{2}\,M_{\oplus}$ \cite[]{Manara:2016}. The pebble flux can be analytically estimated as $\sim100$--300/Myr $M_{\oplus}$ when we assume the disk lifetime is $\sim$1--3 Myr \cite[]{Lambrechts-Johansen:2014}. We fixed the inward pebble mass flux as $\dot{M}_{\rm disk}=10^{2}M_{\oplus}$/Myr , which is consistent with the typical value of the pebble flux used in a previous study \cite[]{Lambrechts:2019}. Since the pebble mass flux is fixed but the radial drift velocity is not, the dust-to-gas ratio, $\Sigma_{\rm p}/\Sigma_{\rm g}$, varies as a function of the Stokes number and the orbital radius,
\begin{align}
\Sigma_{\rm p}/\Sigma_{\rm g}=1.22\times10^{-5}\times\,\Biggr(\frac{1+{\rm St}^{2}}{{\rm St}}\Biggl)\,\Biggr(\frac{a}{1\,\rm{au}}\Biggl)^{1/2},\label{eq:dust-to-gas}
\end{align}
where the pebble surface density is derived from \Equref{eq:M-dot-disk} with the aspect ratio of the disk, $H/a\simeq0.033(a/1\,{\rm au})^{1/4}$, and the typical value of the headwind, $v_{\rm hw}/c_{\rm s}\simeq0.05(a/1\,{\rm au})^{1/4}$, in the MMSN model. 

%---------------------------------------------------------
%---------------------------------------------------------
%---------------------------------------------------------
\section{Results}    \label{sec:result}
\iffigure
 \begin{figure}[htbp]
 \resizebox{\hsize}{!}
 {\includegraphics{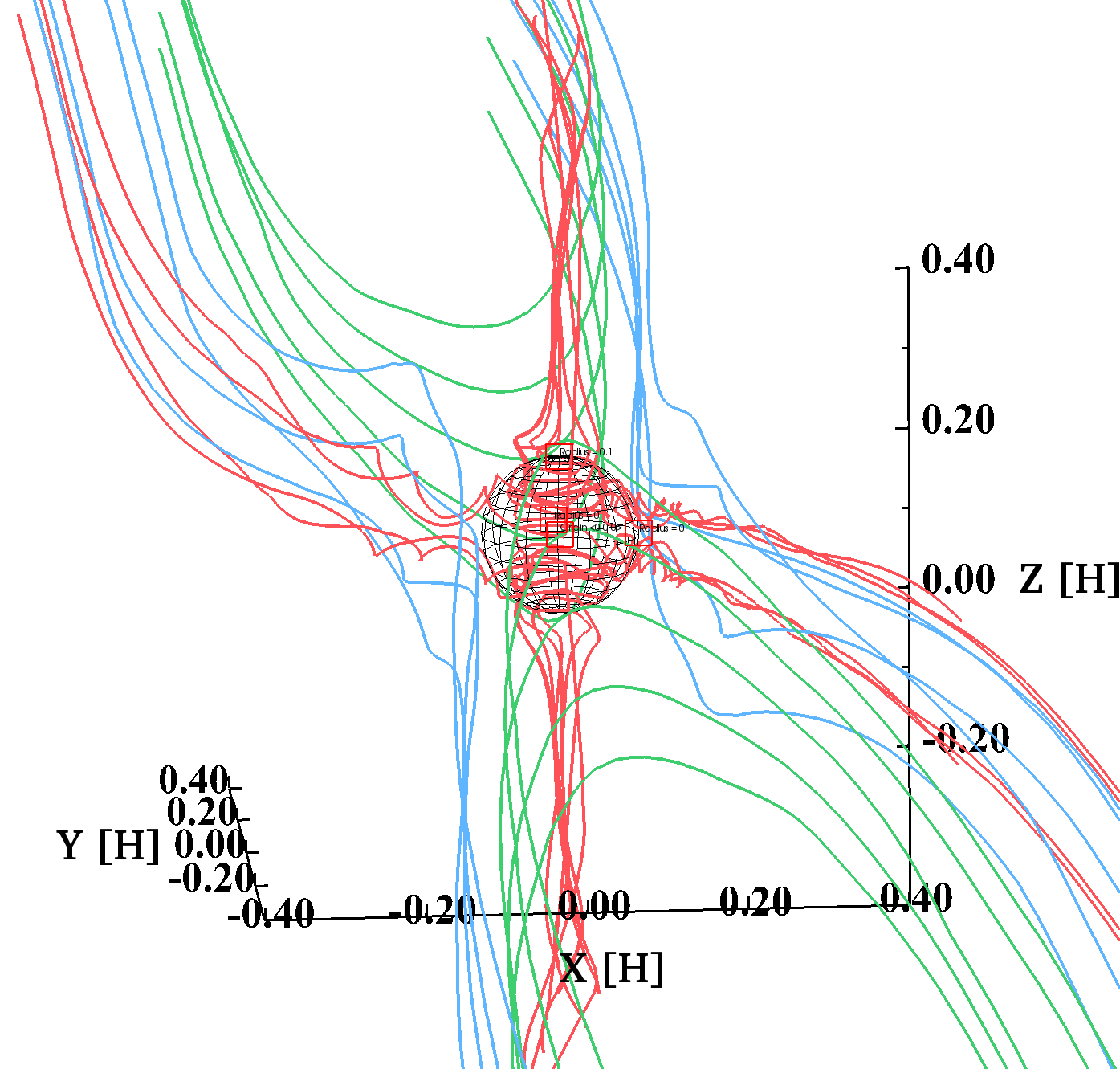}} 
 \caption{Streamlines of 3D planet-induced gas flow around the planet with $m=0.1$ at $t=150$. The red, green, and blue solid lines are the recycling streamlines, the horseshoe streamlines, and the Keplerian shear streamlines, respectively. The sphere is the Bondi sphere of the planet.}
\label{fig:streamline}
\end{figure}
\fi

\iffigure
 \begin{figure*}[htbp]
 \resizebox{\hsize}{!}
 {\includegraphics{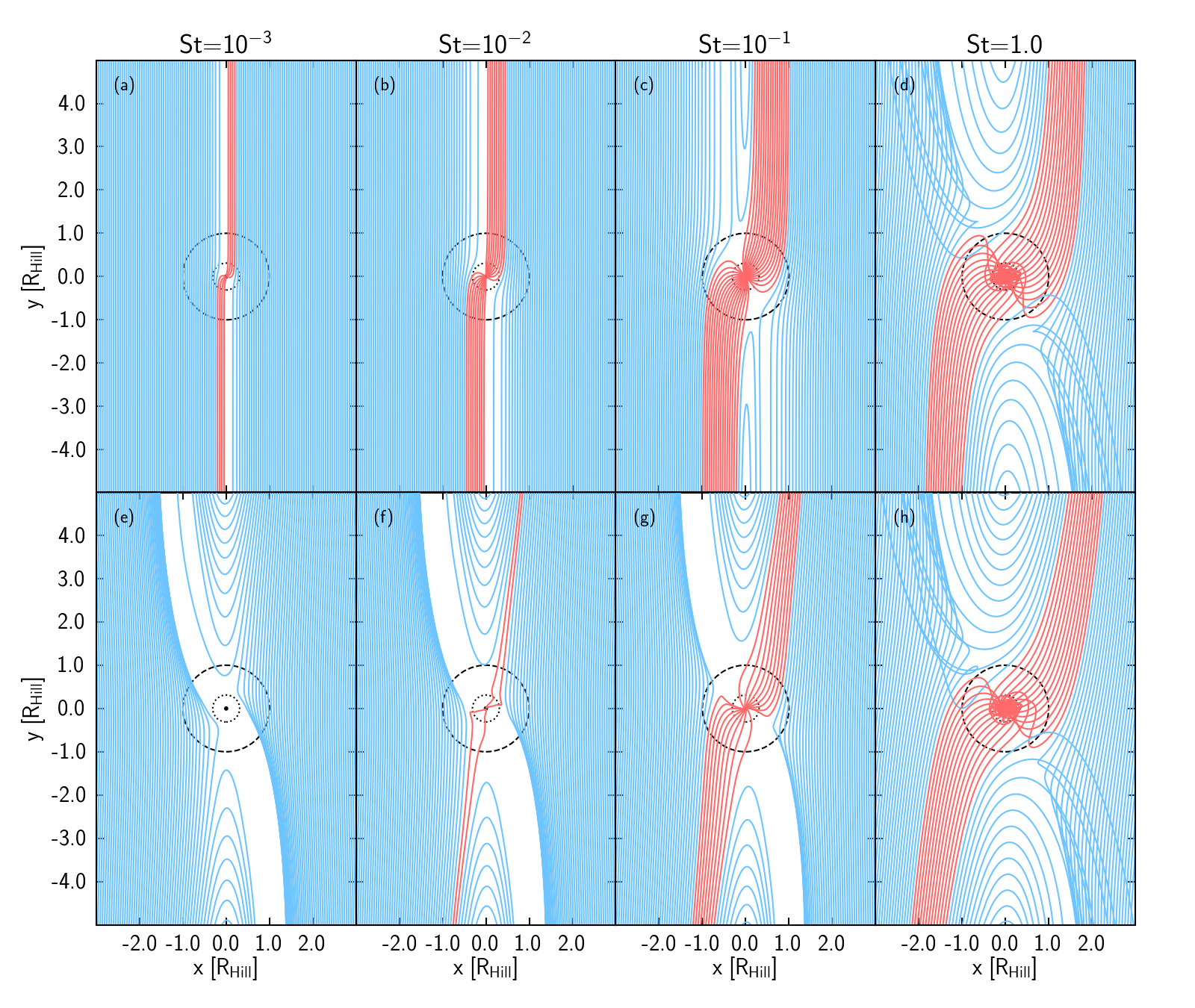}} 
 \caption{Trajectories of pebbles in Shear case (top) and PI-Stokes case (bottom) with different Stokes numbers at midplane around an embedded planet with $m=0.1$. We set $z_{\rm s}=0$ for all cases. The red and blue solid lines correspond to the trajectories of pebbles which accreted and did not accrete onto the planet, respectively. The dashed and dotted circles show the Hill and the Bondi radius of the planet, respectively. The black dots at the center of each panel denote the position of the planet. The interval of pebbles at their initial locations is $0.05$ [H].}
\label{fig:m01-peb-line}
\end{figure*}
\fi

\iffigure
 \begin{figure}[htbp]
 \resizebox{\hsize}{!}
 {\includegraphics{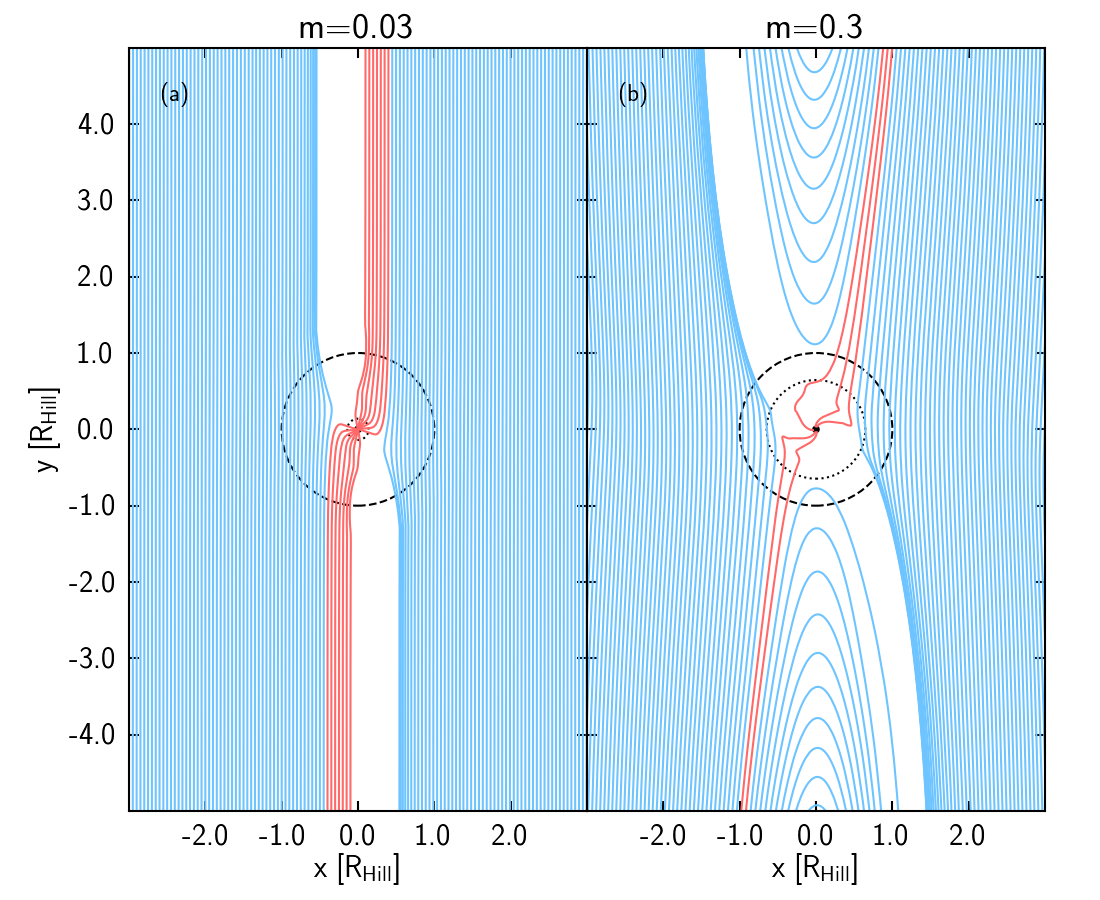}} 
 \caption{Trajectories of pebbles in the PI-Stokes case with ${\rm St}=10^{-2}$ at the midplane around the embedded planets with $m=0.03$ (left panel) and 0.3 (right panel). We set $z_{\rm s}=0$. The red and blue solid lines correspond to the trajectories of pebbles which accreted and did not accrete onto the planet, respectively. The dashed and dotted circles show the Hill and the Bondi radius of the planet, respectively. The sizes of the Hill radius are 0.22 [H] for $m=0.03$ and 0.46 [H] for $m=0.3$. The black dots at the center of each panel denote the position of the planet. The interval of pebbles at their initial locations is $0.05$ [H].}
\label{fig:m003-03-peb-line}
\end{figure}
\fi

\iffigure
 \begin{figure}[htbp]
 \resizebox{\hsize}{!}
 {\includegraphics{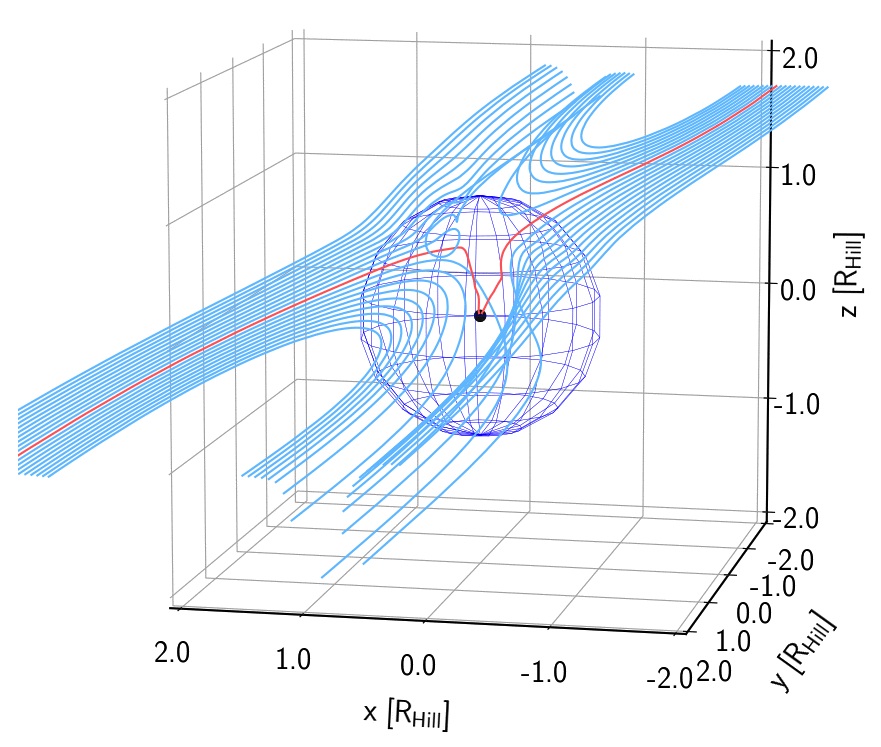}} 
 \caption{Trajectories of pebbles with ${\rm St}=10^{-2}$ around an embedded planet with $m=0.1$ in the PI-Stokes case. The height of pebbles' initial position is $z_{\rm s}=0.7R_{\rm Hill}$. The red and blue solid lines correspond to the trajectories of pebbles which accreted and did not accrete onto the planet, respectively. The black dot denotes the position of the planet. The sphere around the planet represents the Hill radius of the planet. We only plot the trajectories within the region where $r<10R_{\rm Hill}$. The interval of pebbles at their initial locations is $0.05$ [H].}
\label{fig:peb-line-3D}
\end{figure}
 \fi

\iffigure
 \begin{figure}[htbp]
 \resizebox{\hsize}{!}
 {\includegraphics{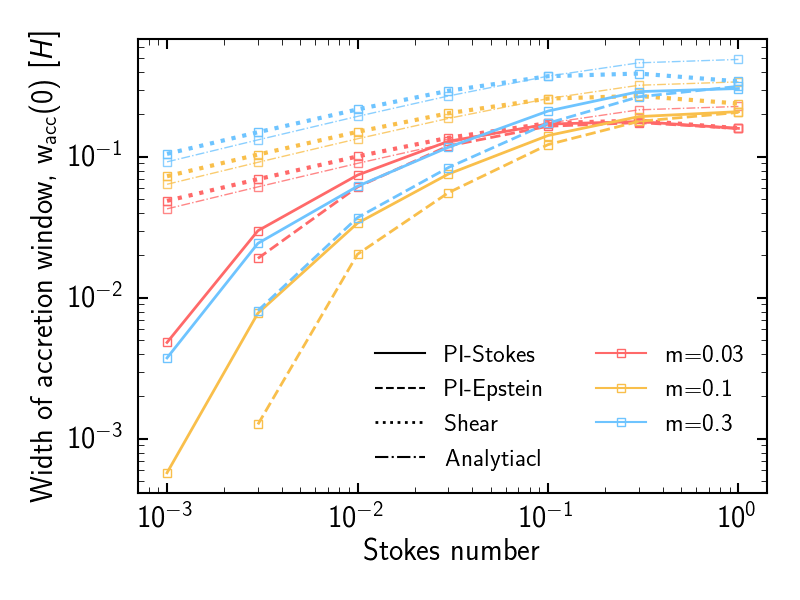}} 
 \caption{Width of accretion window in the midplane, $w_{\rm acc}(0)$, as a function of the Stokes number in the PI-Stokes case (solid lines), PI-Epstein case (dashed-lines), and the Shear case (dotted lines). The dashed-dotted lines correspond to the analytical estimation for the Shear case expressed by \Equref{eq:b-sh}. Colors indicate the mass of the planet: $m=0.03$ (red), $m=0.1$ (yellow), and $m=0.3$ (blue).}
\label{fig:impact-parameter}
\end{figure}
\fi

\iffigure
 \begin{figure*}[htbp]
 \resizebox{\hsize}{!}
 {\includegraphics{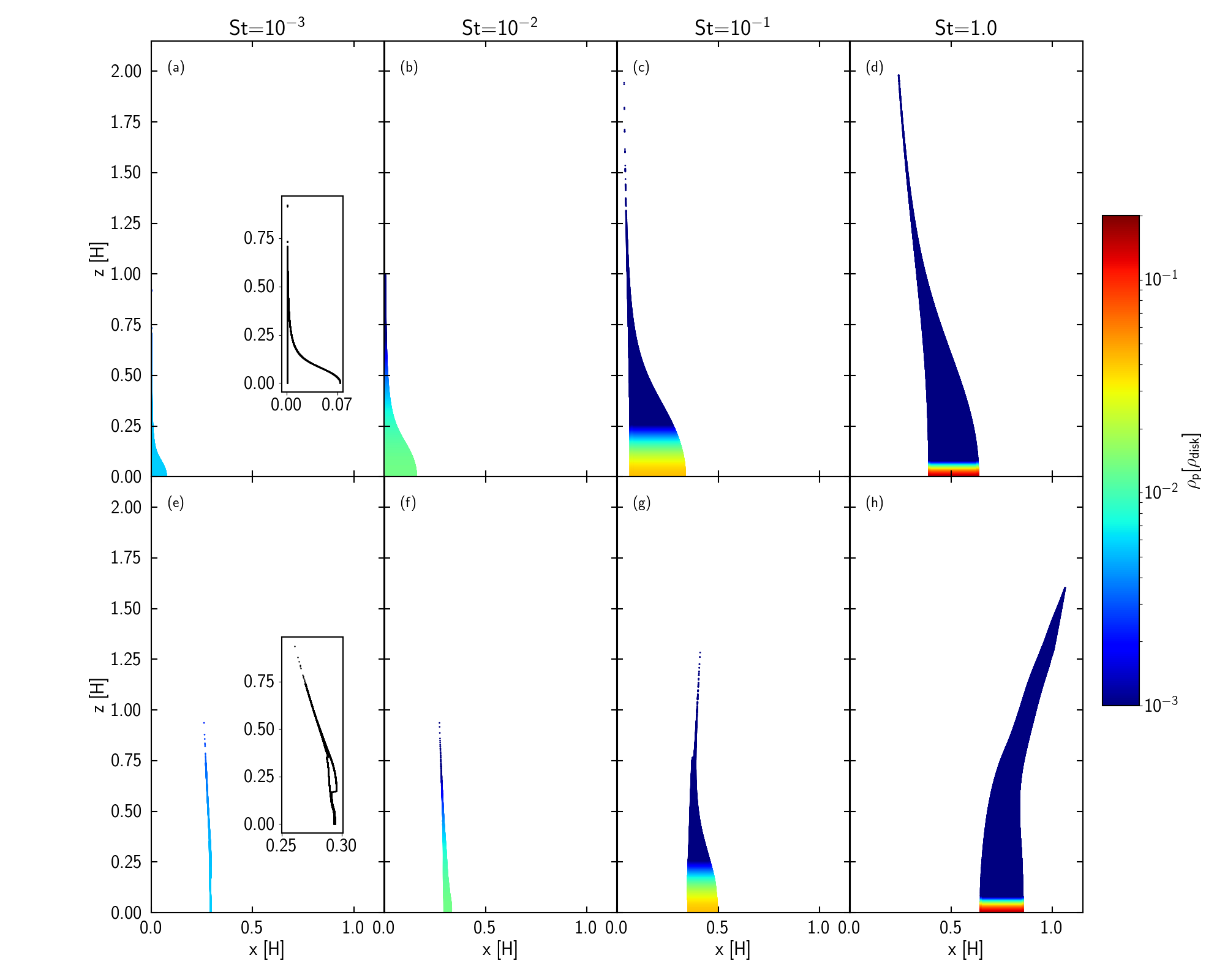}} 
 \caption{Accretion cross section with the different Stokes numbers for the planet with $m=0.1$ in the Shear case (top) and the PI-Stokes case (bottom). We assumed $\alpha=10^{-3}$ and the dust-to-gas ratio is equal to $10^{-2}$. Color represents the density of the pebbles expressed by \Equref{eq:pebble-density} normalized by the gas density. The two panels in panels a and e show the enlarged outlines of accretion cross sections. We note that the color contour is saturated for the $\rho_{\rm p}\lesssim10^{-3}$.}
\label{fig:cross-section}
\end{figure*}
\fi

\iffigure
 \begin{figure}[htbp]
 \resizebox{\hsize}{!}
 {\includegraphics{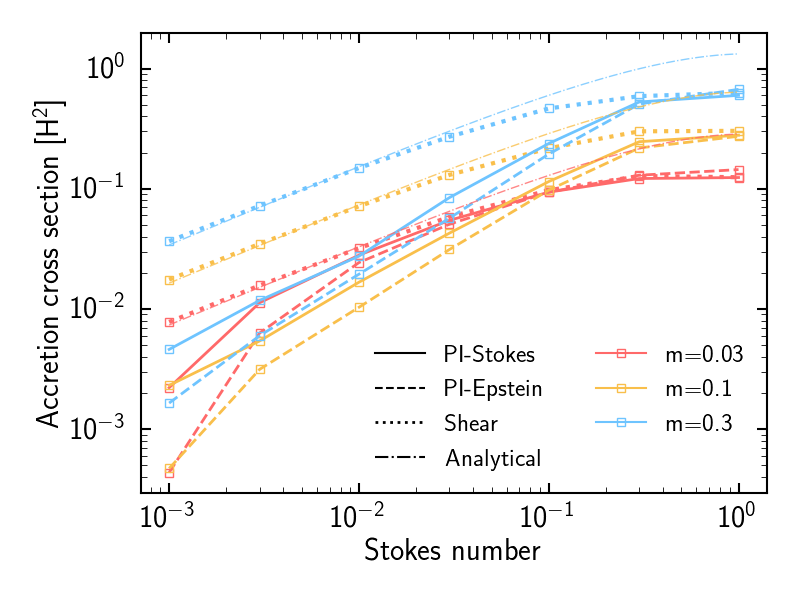}} 
 \caption{Accretion cross section as a function of the Stokes number as a function of the Stokes number in the PI-Stokes case (solid lines), PI-Epstein case (dashed-lines), and the Shear case (dotted lines). The dashed-dotted lines correspond to the analytical estimation for the Shear case expressed by \Equref{eq:acc}. Colors indicate the mass of the planet: $m=0.03$ (red), $m=0.1$ (yellow), and $m=0.3$ (blue).}
\label{fig:integrate-cross-section}
\end{figure}
\fi

\iffigure
 \begin{figure}[htbp]
 \resizebox{\hsize}{!}
 {\includegraphics{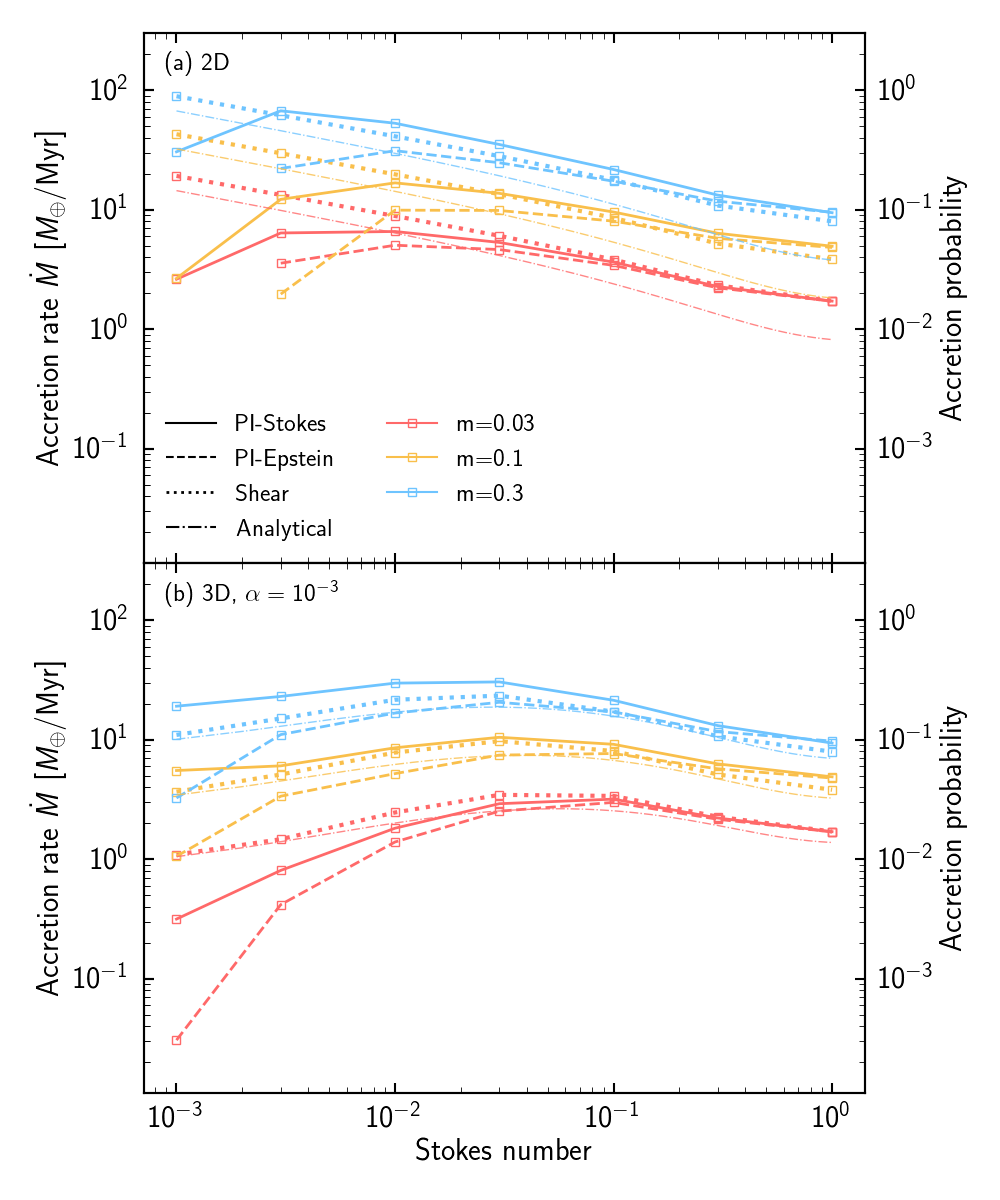}} 
 \caption{Accretion rate (left vertical axis) and probability (right vertical axis)as a function of the Stokes number in the PI-Stokes case (solid lines), PI-Epstein case (dashed-lines), and the Shear case (dotted lines). The dashed-dotted lines correspond to the analytical estimation for the Shear case expressed by Eqs. (\ref{eq:M-dot-2D}) and (\ref{eq:M-dot-3D}). \textit{Top}: 2D case. \textit{Bottom}: 3D case ($\alpha=10^{-3}$). Colors indicate the mass of the planet: $m=0.03$ (red), $m=0.1$ (yellow), and $m=0.3$ (blue).}
\label{fig:M-dot}
\end{figure}
\fi

\iffigure
 \begin{figure}[htbp]
 \resizebox{\hsize}{!}
 {\includegraphics{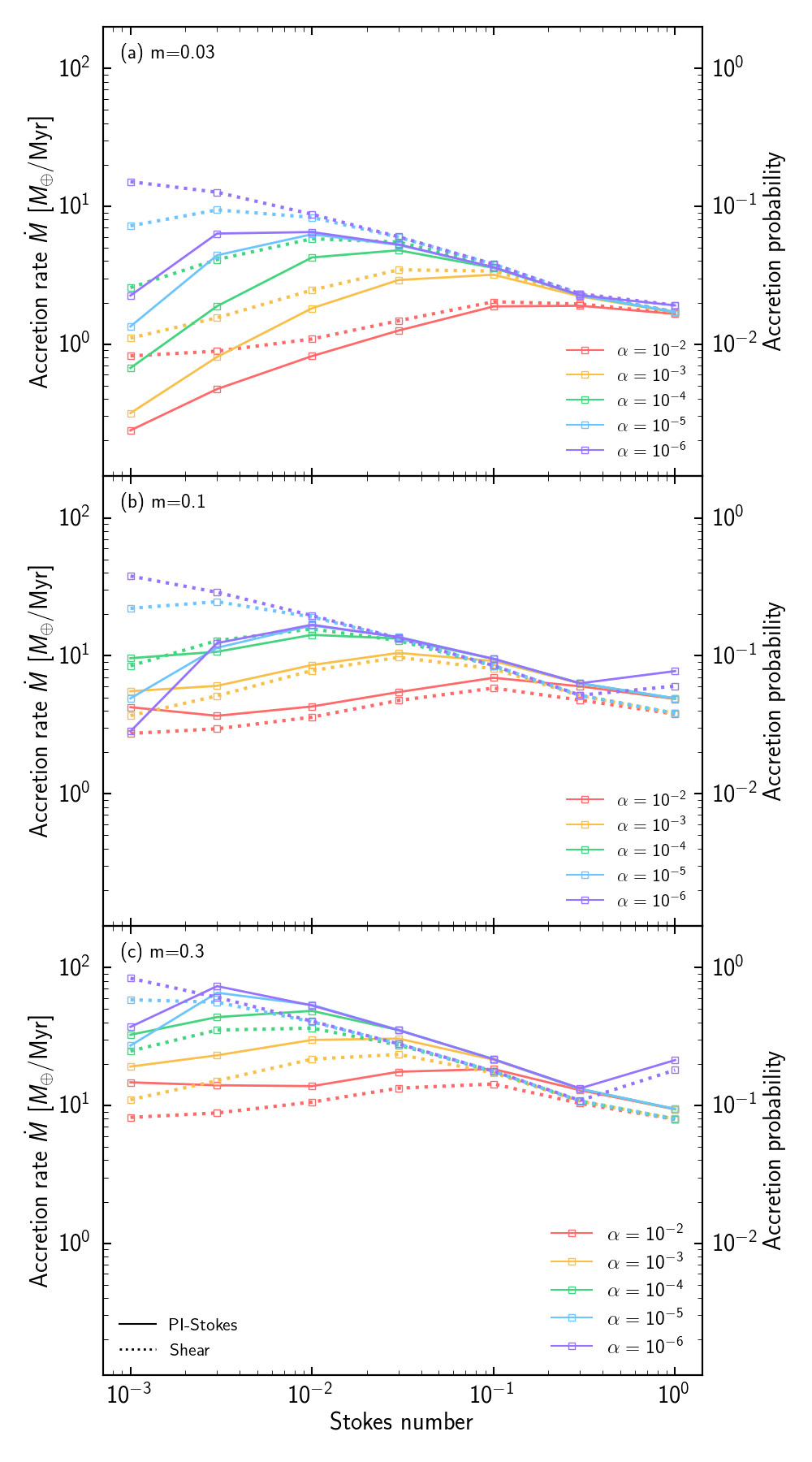}} 
 \caption{Dependence of accretion rate (left vertical axis) and accretion probability (right vertical axis) on the turbulent parameter, $\alpha$. Panels (a), (b), and (c) show the results of the planet with $m=0.03, 0.1$, and 0.3 in the PI-Stokes case (solid lines) and Shear case (dotted lines).}
\label{fig:M-dot-alpha}
\end{figure}
\fi

\iffigure
 \begin{figure}[htbp]
 \resizebox{\hsize}{!}
 {\includegraphics{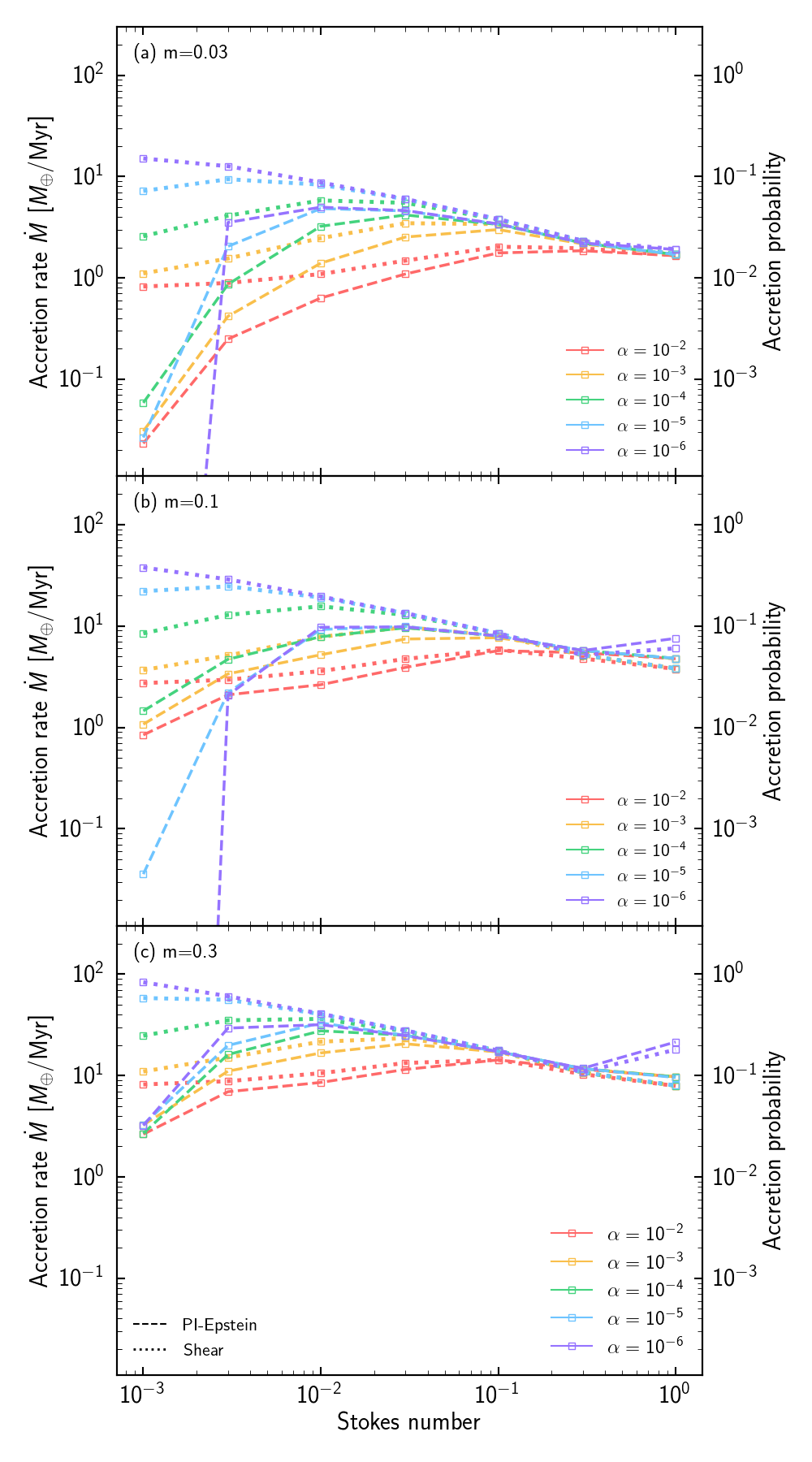}} 
 \caption{Same as \Figref{fig:M-dot}, but the dashed lines show the accretion probability in the PI-Epstein case.}
\label{fig:M-dot-alpha-Ep}
\end{figure}
\fi

\iffigure
 \begin{figure*}[htbp]
 \resizebox{\hsize}{!}
 {\includegraphics{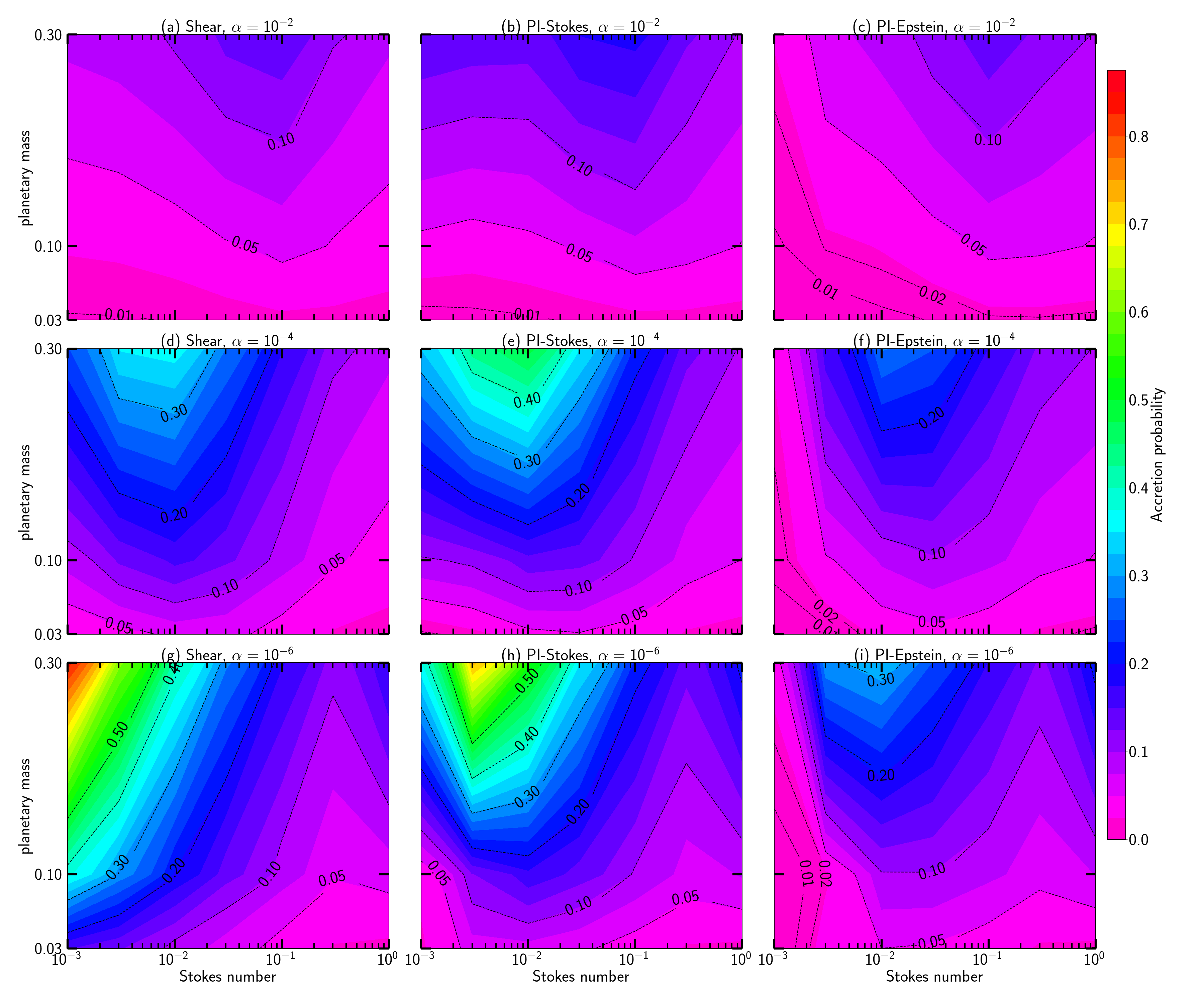}} 
 \caption{Accretion probability as a function of the planetary mass and the Stokes number for the turbulence parameters $\alpha=10^{-6},\,10^{-4},$ and $10^{-2}$ (bottom to top) in the Shear case (left column), the PI-Stokes case (middle column), and the PI-Epstein case (right column). The contours represent the accretion probabilities. }
\label{fig:Contour}
\end{figure*}
\fi
%%%%%%%%%%%%%%%%%%%%%%%%%%%%%%%%
\subsection{Results overview}
The main subject of this study is to clarify the influence of the planet-induced gas flow on pebble accretion. In sect. \ref{sec:result1}, we show the characteristic 3D structure of the planet-induced gas flow field obtained by 3D hydrodynamical simulations. In Sect. \ref{sec:result2},  we show the results of orbital calculations both in 2D and in 3D. Section \ref{sec:result3} shows the dependence of the accretion probability of pebbles on the planetary mass and the Stokes number.
%%%%%%%%%%%%%%%%%%%%%%%%%%%%%%%%%%%%%%%%%%%%
\subsection{3D Planet-induced gas flow}\label{sec:result1}
The gas flow around an embedded planet is perturbed by the planet, which forms the characteristic 3D structure of the flow field \cite[]{Ormel:2015b,Fung:2015,Cimerman:2017,Lambrechts:2017,Kurokawa:2018,Kuwahara:2019,Bethune:2019}. The fundamental features of the flow field are as follows: (1) Gas from the disk enters the Bondi or Hill sphere at high latitudes (inflow) and exits through the midplane region of the disk (outflow) (the red lines of \Figref{fig:streamline}). (2) The horseshoe flow exists in the anterior-posterior direction of the orbital path of the planet (the green lines of \Figref{fig:streamline}). The horseshoe streamlines have a columnar structure in the vertical direction. (3) The Keplerian shear flow extends inside and outside the orbit of the planet (the blue lines of \Figref{fig:streamline}). 

The above-mentioned three features are commonly found both in the isothermal and in the non-isothermal simulations. However, an isolated envelope emerges around the planet in the non-isothermal case. The inner part of the envelope, whose size is $\sim0.5R_{\rm Bondi}$, is isolated from the recycling flow of the gas. The envelope has a slightly higher temperature than that of the disk gas due to compression, but we do not consider the thermal effect (e.g., evaporation) on pebble accretion. The profile of the gas flow in terms of the velocity, the density, the temperature, and the 3D structure does not change after several tens of orbits, which means that the flow field reaches a steady state. The detailed description and the differences between isothermal and non-isothermal 3D gas flow have been summarized in \cite{Kurokawa:2018} and \cite{Kuwahara:2019}. In the following sections, we investigate the influence of the non-isothermal planet-induced gas flow on pebble accretion.

\subsection{Orbital calculations}\label{sec:result2}
\subsubsection{Pebble accretion in 2D}\label{sec:PA2D}
Firstly we focus on the 2D limit of pebble accretion in the Stokes regime, that is all of the pebbles settled in the midplane of the disk and the Stokes number of pebbles does not depend on the gas density. Figure \ref{fig:m01-peb-line} shows the trajectories of pebbles at the midplane of the disk. The width of the accretion window becomes narrower in the PI-Stokes case than in the Shear case, in particular for the smaller Stokes numbers, ${\rm St}\lesssim0.1$ (Figs. \ref{fig:m01-peb-line}e--g). In the Shear case, the pebbles that successfully accrete onto the planet originated in the vicinity of the planetary orbit, $x=0$. In the PI-Stokes case, however, most pebbles coming from this region move away from the planet along the horseshoe flows. This trend is consistent with the result of a previous study \cite[]{Popovas:2018a}. The pebbles coming from a window between the horseshoe and the shear regions can accrete onto the planet. We found voids in the upper left and lower right regions of the planet when the pebbles are small (Figs. \ref{fig:m01-peb-line}e--g). The outflow speed of the gas at the midplane region of the disk is given by $\sqrt{3}/2\ mc_{\rm s}$ for the range of planetary mass considered, which influences small pebbles when ${\rm St}\lesssim\sqrt{m}$ as predicted by \cite{Kuwahara:2019}. The midplane outflow and resulting voids in pebble trajectories are not found in 2D simulations \cite[][their Fig. 12]{Ormel:2013}, but appear in the 3D ones.

For the larger Stokes number, ${\rm St}=1$, we obtained similar results both in the Shear and in the PI-Stokes cases (Figs. \ref{fig:m01-peb-line}d and h). When the Stokes number becomes large, that is the stopping time of the pebble increases, pebbles are loosely coupled with the gas. Though the disk gas is perturbed by the planet, its influences on the larger pebbles become weak. Nevertheless, the motion of large pebbles is partially affected by the planet-induced gas flow on a large scale. We found that the shape of the accretion cross section of pebbles in the planet-induced gas flow differs from that in the unperturbed gas flow (see Sect. \ref{sec:3D} and \Figref{fig:cross-section}). In the PI-Epstein case, the shape of trajectories of pebbles does  not differ significantly from that in the PI-Stokes case when we plot the trajectories with the same intervals of \Figref{fig:m01-peb-line}, 0.05 [H]. However, the width of the accretion window, the accretion cross section, and the accretion probability in the PI-Epstein case do not match with those in the PI-Stokes case (see Sects. \ref{sec:re3} and \ref{sec:re4}).

The dependence on the planetary mass in the PI-Stokes case is shown in \Figref{fig:m003-03-peb-line}. In our previous study, we have found that the planet chiefly perturbs the surrounding disk gas in the typical scale of the smaller of the Bondi and Hill radii, and the outflow speed increases with the planetary mass \cite[]{Kuwahara:2019}. In both Figs. \ref{fig:m003-03-peb-line}a and b, by perturbed region can be scaled by the Bondi radius, but their sizes are different by an order of magnitude. When the planetary mass becomes larger, the perturbed region and the outflow speed become larger. Therefore, one can see the significant difference in Figs. \ref{fig:m003-03-peb-line}a and b shown with the same scale. Whereas the pebbles with ${\rm St}=10^{-2}$ are hardly influenced by the perturbation in the gas flow induced by the planet when $m=0.03$, the same-sized pebbles are highly influenced when $m=0.3$.

\subsubsection{Pebble accretion in 3D}\label{sec:3D}
Next we focus on the 3D behavior of pebble accretion in the Stokes regime. Figure \ref{fig:peb-line-3D} shows the 3D trajectories of pebbles with ${\rm St}=10^{-2}$ around an embedded planet with $m=0.1$ in the PI-Stokes case. The behavior of pebble accretion does not differ significantly from that in the midplane (\Figref{fig:m01-peb-line}f). Since the horseshoe streamlines extend in the vertical direction keeping its configurations (\Figref{fig:streamline}), pebbles coming from the vicinity of planetary orbit move away from the planet and do not accrete onto it. The pebbles coming from the narrow region between the horseshoe and the shear regions sharply descend toward the planet with terminal velocity when they approach the planet from the polar region. This downward motion is due to the gravity of the planet and the advection due to the gas inflow (\Figref{fig:streamline}). Disk gas enters the Bondi sphere from the polar region and from the altitude where the planet gravity dominates ($z\sim R_{\rm Bondi}$). The inflow has the velocity component of $-z$-direction, which causes rapid changes in the motion of the  pebbles. The outflow of the gas also deflects the trajectories of pebbles and make the voids in the second and fourth quadrants of $x$-$y$ plane because the outflows extend in the vertical direction \cite[]{Kuwahara:2019}. The vertical scale of the outflow is $\sim R_{\rm Bondi}$ in the non-isothermal case. 

%%%%%%%%%%%%%%%%%%%%%%%%%%%%%%
\subsection{Accretion probability of pebbles}\label{sec:result3}
\subsubsection{Width of accretion window and accretion cross section}\label{sec:re3}

%%%%%%%%%%%%%%%%%%%%%%%%%%%%%%%%%

Figure \ref{fig:impact-parameter} shows the changes of the width of the accretion window in the midplane region as a function of the Stokes number for different planetary masses $m=0.03$, 0.1, and 0.3. In the Shear case, the numerically calculated width of the accretion window is almost consistent with the analytical estimation of the width of the accretion window (\Equref{eq:b-sh}). The width of the accretion window in the unperturbed shear flow follows a power law with index 1/3. 

In the PI-Stokes case, since the gas flow effects are weaker for the larger pebbles, the widths of accretion windows almost agree with those in the Shear case when ${\rm St\gtrsim0.01}$ (for $m=0.03$) or ${\rm St}\gtrsim0.1$ (for $m=0.1$, 0.3). On the other hand, the widths of accretion windows deviate from those in the Shear case when the Stokes number is small, ${\rm St}\lesssim10^{-2}$--$10^{-1}$. In the smallest Stokes number, ${\rm St}=10^{-3}$, the width of the accretion window in the PI-Stokes case decreases by one or two orders of magnitude compared to that in the Shear case. 
 
The reduction of the widths of accretion windows at the midplane becomes more prominent in the PI-Epstein case. This is because, in the Epstein regime, the Stokes number is proportional to the reciprocal of gas density. Since the gas density is higher around the planet due to its gravity, the effective Stokes number decreases as the pebble approaches the planet. In particular, pebbles with ${\rm St}=10^{-3}$ do not accrete onto the planet with the spatial intervals of $0.01b_{x}$ in the $x$-direction. This means that the width of the accretion window is smaller than $w_{\rm acc}(0)\lesssim10^{-4}$ for ${\rm St}=10^{-3}$. We truncate the width of the accretion window in PI-Epstein case at ${\rm St}=3\times10^{-3}$ in \Figref{fig:impact-parameter}.

We plotted the accretion cross section of pebbles both in the Shear case and in the PI-Stokes case for a planet with $m=0.1$ (\Figref{fig:cross-section}). In both cases, the accretion cross section decreases with the Stokes numbers.

Firstly, we consider the results for the smaller Stokes numbers, ${\rm St}\leq0.1$ (Figs. \ref{fig:cross-section}a--c, e--g). In the Shear case, since the pebbles approach the planet almost linearly (\Figref{fig:m01-peb-line}), the projected accretion cross sections are located near the planetary orbit. The value of $x_{\rm min}(z)$ hardly changes, but the $x_{\rm max}(z)$ value decreases with height. Since the gravity of the planet acting on the pebbles becomes weaker at high altitudes, $x_{\rm max}(z)$ should decrease with height. 

In the PI-Stokes case, the accretion cross section shifts to the right as a whole, and its shape becomes narrower compared to that in the unperturbed shear flow (Figs.
 \ref{fig:cross-section}e--g). This is due to the horseshoe flow, which lies across the orbit of the planet and has a columnar structure in the vertical direction. A notable result can be seen in the zoomed view in \Figref{fig:cross-section}e. We found an accretion window above the midplane region. The width of the accretion window broadens at high altitudes, $z\sim0.2$--0.3 $[H]$. The emergence of this window related to the vertical scale of the outflow of the gas. In the midplane region, the  pebbles with ${\rm St}=10^{-3}$ are inhibited from accreting onto the planet by the planet-induced outflow. However, the pebbles approaching from an altitude higher than the vertical extent of outflow ($z\sim R_{\rm Bondi}$) can accrete onto the planet. Accretion windows at high latitudes above the outflow region for small pebbles were also found  for other planet masses.

For the larger Stokes number, ${\rm St}=1.0$, the accretion cross section shifts to the right  in both the Shear case and the PI-Stokes case (Figs. \ref{fig:cross-section}d and h). When the Stokes number becomes larger, the three-body effect becomes more important than the effect of the gas. Thus the behavior of the pebble is similar in both the Shear and the PI-Stokes cases in the region close to the planet near the midplane of the disk. However, at high altitudes, $z\gtrsim0.6$ [H], we found the opposite trend between Figs. \ref{fig:cross-section}d and h. In the PI-Stokes case, the tip of the accretion cross section turns to the right in contrast to the result in the Shear case (\Figref{fig:cross-section}d). 

Figure \ref{fig:integrate-cross-section} shows the differences between the integrated accretion cross section in the Shear, PI-Stokes, and PI-Epstein cases, and the dependence on the planetary mass. In the Shear case, the numerically calculated accretion cross section matches the analytical estimation (\Equref{eq:acc}). The accretion cross section in the unperturbed shear flow follows a power law with index 2/3.

In the PI-Stokes case, the accretion cross sections agree with those in the Shear case when ${\rm St\gtrsim0.01}$ (for $m=0.03$) or ${\rm St}\gtrsim0.3$ (for $m=0.1$ and 0.3). On the other hand, the accretion cross sections deviate from those in the Shear case when the Stokes number is small, ${\rm St}\lesssim3\times10^{-3}$--$10^{-1}$. For the pebbles with ${\rm St}=10^{-3}$, the accretion cross section in the PI-Stokes case decreases by an order of magnitude compared to that in the Shear case. 

Similarly to the 2D case (\Figref{fig:impact-parameter}), the reduction of the accretion cross section becomes more prominent in the PI-Epstein case (\Figref{fig:integrate-cross-section}). We note that, unlike \Figref{fig:impact-parameter}, we can compute the accretion cross section for the pebbles with ${\rm St}=10^{-3}$ because $w_{\rm acc}(z)$ takes a non-zero value above the outflow region.

\subsubsection{Accretion probability}\label{sec:re4}
Figure \ref{fig:M-dot} shows the accretion rate and the accretion probability as a function of the Stokes number for the different planetary masses. In the Shear case, the numerical results are consistent with the analytical estimations both in 2D and 3D (Eqs. (\ref{eq:M-dot-2D}) and (\ref{eq:M-dot-3D})). The accretion rate (probability) increases as the planetary mass increases, but decreases as the Stokes number increases in 2D \cite[]{Liu:2018}. In 3D, however, the accretion probability has a peak at ${\rm St}\sim0.03$ \cite[]{Ormel-note,Ormel:2018}. 

In the PI-Stokes case, the situation changed. In the 2D case, the accretion probabilities either match or are slightly larger than that in the Shear case when ${\rm St}\gtrsim3\times10^{-3}$--$10^{-2}$, but deviate from it when ${\rm St}$ become smaller than the preceding value (\Figref{fig:M-dot}a). Suppression of pebble accretion becomes significant when the Stokes number is small in the 2D limit. However, the coincidence in accretion probabilities between unperturbed and perturbed flow cases for ${\rm St}\gtrsim3\times10^{-3}$--$10^{-2}$ is counterintuitive, in particular for $m=0.1$ and 0.3 because the width of the accretion window
 decreases significantly in the planet-induced gas flow when ${\rm St}\lesssim10^{-1}$ for $m=0.1$ and 0.3 (\Figref{fig:impact-parameter}). This phenomenon can be understood as follows: in the Shear case, the accretion window lies near the planetary orbit, where the relative velocity between pebbles and the planet is low because the velocity is given by $\bm{v}_{\rm p, \infty}=-3/2x\bm{e}_{y}$. In the PI-Stokes case, the accretion cross section shifted to the right (\Figref{fig:cross-section}), where the relative velocity is large. In the midplane region, the offset of the reduction of the accretion cross section by the increase of relative velocity led to the accretion probability in the perturbed flow case being almost equal to that in the shear flow for ${\rm St}\gtrsim3\times10^{-3}$--$10^{-2}$.

In the 3D case where we assumed $\alpha=10^{-3}$ as a nominal value, we found the accretion probability in the PI-Stokes case either matches or is slightly larger than that in the Shear case for the planet with $m=0.1$ and 0.3 (\Figref{fig:M-dot}b). Similarly to the 2D case, the reduction of the accretion cross section and the increase of relative velocity cancel each other out.  Since the width of the accretion window of the pebbles with ${\rm St}=10^{-3}$ increases at high altitudes (\Figref{fig:cross-section}e), the accretion probability in the planet-induced gas flow never falls below that in the unperturbed shear flow. 

The above explanations shall not apply to $m=0.03$, where the accretion probability for the smaller pebbles in the PI-Stokes case is smaller than that in the Shear case. Because the horseshoe width is proportional to the square of the planetary mass \cite[]{Masset:2016}, the shifting rate of the accretion cross section decreases as the planetary mass decreases. As a consequence, when the planetary mass is small, $m=0.03$, the accretion cross section does not shift to the right enough to cancel its reduction.

 In the PI-Epstein case, the reduction of the width of the accretion window and the accretion cross section become more significant than those in the PI-Stokes case for the smaller pebbles (Figs. \ref{fig:impact-parameter} and \ref{fig:integrate-cross-section}). The increase of the relative velocities of pebbles does not fully offset the significant reduction of the width of the accretion window and the accretion cross section. Therefore, the accretion probability in the PI-Epstein case becomes smaller than that in the Shear- and PI-Stokes cases regardless of assumed ${\rm St}$ and $m$. Since the pebbles with ${\rm St}=10^{-3}$ do not accrete onto the planet in the Epstein regime in the 2D case, the accretion probabilities are truncated at ${\rm St}=3\times10^{-3}$ in \Figref{fig:M-dot}a. In the last paragraph of Sect. \ref{sec:PA2D} we expected $w_{\rm acc}(0)\lesssim10^{-4}$ for ${\rm St}=10^{-3}$.  Based on \Equref{eq:M-dot-2D}, we have
 \begin{align}
\dot{M}_{\rm p,2D}&\propto\left(x^{2}_{\rm max}(0)-x^{2}_{\rm min}(0)\right),\nonumber \\
&=\left(x_{\rm max}(0)+x_{\rm min}(0)\right)w_{\rm acc},
\end{align}
where $x_{\rm max}(0)+x_{\rm min}(0)$ is expected to have an order of $\sim10^{-1}$. If the pebbles can accrete onto the planet in the midplane with narrower spatial intervals than those investigated in this study, from these estimates we expect that the accretion probability for the pebbles with ${\rm St}=10^{-3}$ is $P_{\rm acc}\lesssim4\times10^{-4}$. This estimation proved to be valid when $m=0.03$ and 0.1 (see the following paragraph and \Figref{fig:M-dot-alpha-Ep}).

Figures \ref{fig:M-dot-alpha} and \ref{fig:M-dot-alpha-Ep} show the dependence on the turbulent parameter. Each panel shows the accretion probability for a planet with $m=0.03$, 0.1, and 0.3. The accretion probability in the Shear case increases as $\alpha$ decreases and approaches that in the 2D case. When $m=0.03$ (\Figref{fig:M-dot-alpha}a), the accretion probability in the PI-Stokes case also increases as $\alpha$ decreases, and remains smaller than that in the Shear case. However, for $m=0.1$ and 0.3 (Figs. \ref{fig:M-dot-alpha}b and c), the accretion probability for the smaller pebbles (${\rm St}\lesssim10^{-3}$--$10^{-2}$) becomes smaller than that in the Shear case, only when the turbulent parameter falls below $\alpha\lesssim10^{-5}$.

In the PI-Epstein case, the accretion probability for a pebble with ${\rm St}=10^{-3}$ takes a much smaller value, $P_{\rm acc}\lesssim4\times10^{-4}$, in the weak turbulence for a planet with $m=0.03$ and 0.1 as we mentioned before (Figs. \ref{fig:M-dot-alpha-Ep}a and b). The accretion probability for a planet with $m=0.3$ converges at $P_{\rm acc}\sim2\times10^{-2}$ in the range of turbulent parameter used in this study (\Figref{fig:M-dot-alpha-Ep}c).  In the PI-Epstein case, accretion probabilities never exceed that in the Shear case even for strong turbulence, $\alpha=10^{-2}$. 

Figure \ref{fig:Contour} shows the accretion probability as a function of both the planetary mass and the Stokes number for the various turbulence strengths, $\alpha$. Under strong turbulence, the accretion probability is always below $P_{\rm acc}\lesssim0.2$ (Figs. \ref{fig:Contour}a--c). The accretion probability has a peak at ${\rm St}\sim0.1$. As the turbulence strength decreases, the maximum value of the accretion probability increases and the peak shifts to smaller St (Figs. \ref{fig:Contour}d--i). Suppression of pebble accretion due to the planet-induced gas flow becomes prominent for smaller Stokes numbers and smaller turbulence parameters. The peak of accretion probability lies in the upper left (higher $m$ and smaller St) region of \Figref{fig:Contour}g and has $P_{\rm acc}\gtrsim0.5$, whereas the accretion probability remains below $P_{\rm acc}\lesssim0.05$ in the corresponding region of \Figref{fig:Contour}i.
%---------------------------------------------------------
%---------------------------------------------------------
%---------------------------------------------------------

\section{Discussion}    \label{sec:discussion}
\iffigure
 \begin{figure}[htbp]
 \resizebox{\hsize}{!}
 {\includegraphics{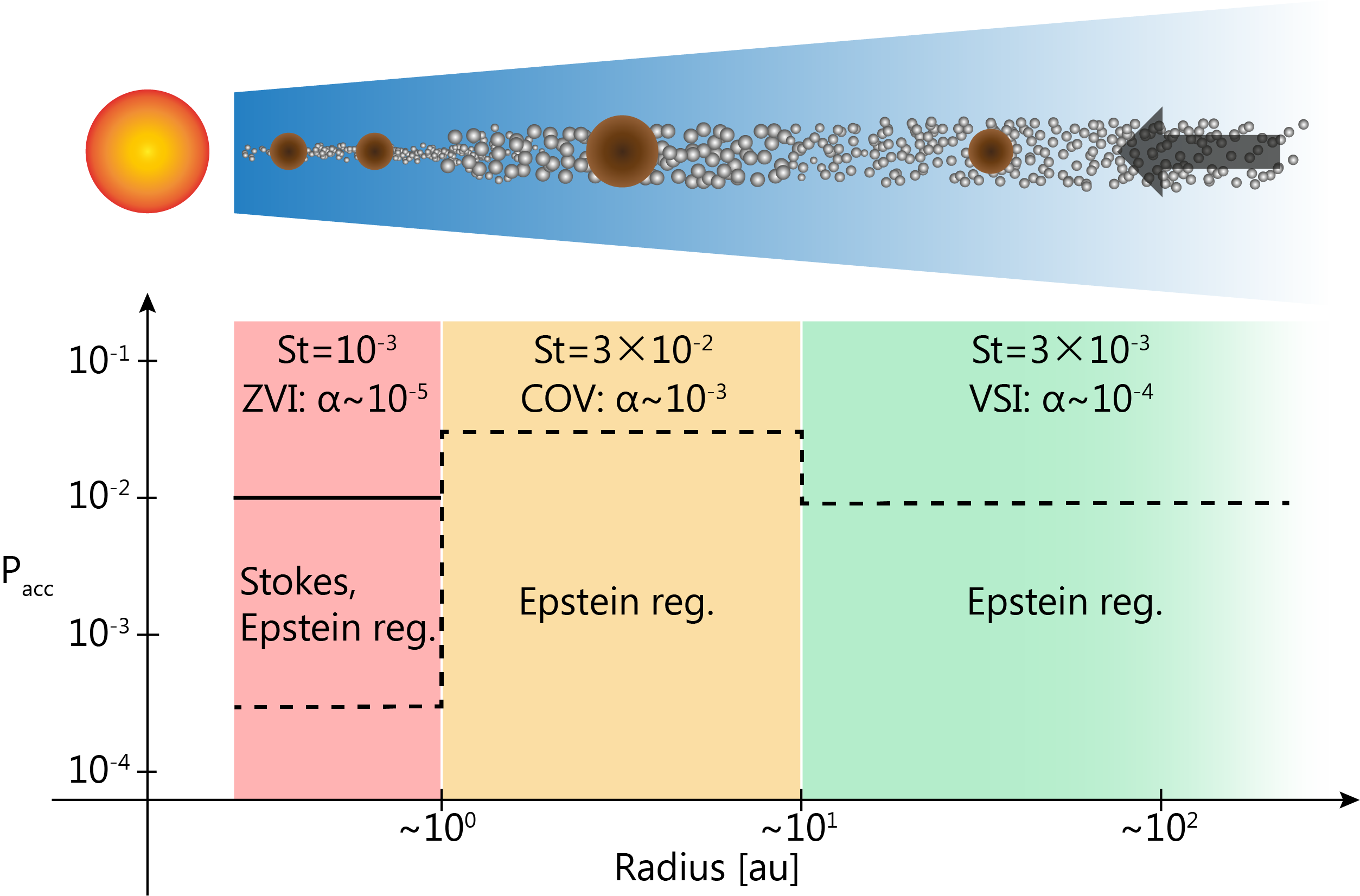}} 
 \caption{Formation scenario of planetary system. \textit{Top:} the schematic picture of the protoplanetary disk. The brown and white filled circles denote the planets and drifting pebbles. \textit{Bottom:} the assumed Stokes numbers, the origins of the turbulence (ZVI, COV, and VSI), and the resulting accretion probability in the Stokes (solid line) and in the Epstein (dashed line) regimes.}
\label{fig:summary}
\end{figure}
\fi
%%%%%%%%%%%%%%%%%%%%%%%%%%%%%%%%%%%
\subsection{Comparison to previous studies}
Recently, \cite{Popovas:2018a} investigated  pebble accretion onto the Mars- to Earth-mass planets ($m=0.008,\,0.04,\,$ and 0.08) in the planet-induced gas flow in the Epstein regime. They performed 3D non-isothermal simulations in the shearing box. After the sufficient relaxation of the fluid motion and density distribution of the gas, they injected millions of tracer particles with different sizes (${\rm St}=3\times10^{-5}$--3). The initial stratification of particles is the same as that of the gas. They measured the accretion rates of a sub-population of particles that initially resided within the Hill sphere. They have reported that the accretion rate increases as the planetary mass increases, and it scales linearly with pebble size regardless of the assumed planetary mass when ${\rm St}=3\times10^{-5}$--$3\times10^{-1}$. In their subsequent study, they concluded that the convective motion within the envelope caused by the accretion heating does not significantly affect the accretion rate \cite[]{Popovas:2018b}. Similarly to their results, the accretion rates increase with the planetary mass (\Figref{fig:M-dot-alpha-Ep}). In our study, however, the slope of the accretion rate changes significantly for the assumed planetary mass and turbulent parameter. The accretion rate does not scale linearly over the same pebble size range as \cite{Popovas:2018a}. The maximum accretion rate in \cite{Popovas:2018a} is $\dot{M}\sim10^{2}$ $M_{\oplus}$/Myr when $m=0.08$ and ${\rm St}\sim1$--3, but the achieved accretion rate in our study becomes smaller by one or two orders of magnitude even for a larger planet, $m=0.3$ (Figs. \ref{fig:M-dot-alpha-Ep}b and c). The higher accretion rate obtained in \cite{Popovas:2018a} may be the result of the absence of turbulent stirring. In contrast to our study, \cite{Popovas:2018a} considered the vertical component of the tidal force, $-z\Omega\bm{e}_{z}$, so that the particles settle to the midplane with the settling timescale, $({\rm St}\Omega)^{-1}$, which may lead to a high accretion rate for the larger pebbles. We discuss the effect of the turbulence on the accretion probability in Sect. \ref{sec:turb}.

In our previous study, we found that the speed of outflow of gas at the midplane region can be expressed by $u_{\rm out}=\sqrt{3/2}mc_{\rm s}$ \cite[]{Kuwahara:2019}. Comparing the outflow speed to the terminal velocity of pebbles, we found that planetary mass with the potential to affect the accretion can be written as $m\gtrsim\sqrt{{\rm St}}$. As the mass of the planet increases, the outflow becomes fast and would start to prevent solid materials from accreting onto the core. As shown in Figs. \ref{fig:impact-parameter}, \ref{fig:cross-section}, and \ref{fig:integrate-cross-section}, the reduction of the accretion cross section becomes significant for the pebbles with ${\rm St}\lesssim m^{2}$, which is consistent with the prediction in \cite{Kuwahara:2019}. However, the outflow would not completely prevent pebble accretion. Most pebbles approach the planet from the first and third quadrants of the $x$-$y$ plane. In contrast, the outflow occurs dominantly in the second and fourth quadrants of the $x$-$y$ plane. 

\subsection{Topologies of planet-induced gas flow}
\subsubsection{Width of horseshoe flow}
\label{sec:HS}
The shift of the accretion cross section is controlled by the width of the horseshoe region given by \cite{Masset:2016}
\begin{align}
w_{\rm HS}=1.05\sqrt{m}.\label{eq:horseshoe-width}
\end{align}
Equation (\ref{eq:horseshoe-width}) agrees with the horseshoe width of our hydrodynamical simulations for \texttt{m01} and \texttt{m03} runs. In \texttt{m003} run, \Equref{eq:horseshoe-width} overestimated the horseshoe width. This overestimation of the horseshoe width for a lower mass planet ($m\leq0.03$) could be seen in the previous studies \cite[]{Ormel:2015b,Kurokawa:2018,Kuwahara:2019}, but if $w_{\rm HS}$ continues to decrease as the planetary mass decreases, the accretion cross section would hardly shift. Thus, whether the accretion probability changes depends on whether the outflow barrier works for pebbles in the vicinity of the planet. However, the outflow speed decreases with the planetary mass, and it becomes difficult for the outflow to suppress pebble accretion when the planetary mass is small \cite[]{Kuwahara:2019}. Assuming that the lower limit of the Stokes number is ${\rm St}=10^{-3}$, according to our previous study, we would expect that only a planet larger than $m\geq0.03$ changes the accretion probability \cite[]{Kuwahara:2019}. Pebble accretion in the planet-induced gas flow for the lower mass planets may not be much different from that in the unperturbed shear flow, as long as we focus on the shear regime of pebble accretion. When the planetary mass is much smaller than what is considered in this study, the effect of the headwind becomes more important (see the following section).

\subsubsection{Effect of headwind}
Pebble accretion probability depends on the topologies of the planet-induced gas flow. In this study, we did not consider the sub-Kepler motion of the gas as we focused on the shear regime. Given the non-zero $\eta$ (\Equref{eq:eta}), the flow topology changes \cite[]{Ormel:2013,Ormel:2015a,Ormel:2015b,Kurokawa:2018}. The flow topology is nearly axisymmetric with respect to $x=0$ when we assume $\eta=0$, but the symmetry breaks down when $\eta\neq0$. The horseshoe streamlines shift to within the planetary orbit. Such asymmetric flow structure may affect the accretion probability of pebbles. The shift of the horseshoe may enhance the accretion of pebbles coming from the front of a planet's orbit, whereas the relative velocity of accreted pebbles (Sect. \ref{sec:result3}) may be reduced. Therefore, the net effect of the headwind is not obvious. Considering the headwind regime is beyond the scope of this study, but it would be important for understanding pebble accretion onto a core smaller than the transition mass (\Equref{eq:transition}). 

\subsubsection{Effect of turbulence}\label{sec:turb}
The strong turbulent viscosity also changes the flow topology. \cite{Fung:2015} have performed 3D viscous hydrodynamical simulations. They have found that the outward gas flow streamline vanished, then inflowing streams entering the Bondi sphere emerged between the horseshoe and shear streamlines in the first and third quadrants in the $x$-$y$ plane in the midplane of the disk when $\alpha=10^{-2}$. This inflow may promote the accretion of pebbles and increase the accretion probability. 

The effects of the turbulence on the accretion probability are investigated in previous studies. \cite{Xu:2017} investigated the effect of turbulent stirring on pebble accretion. They performed three types of local 3D hydrodynamical simulations under the shearing-sheet approximation: the ideal magnetohydrodynamical (MHD) simulation ($\alpha=4.5\times10^{-2}$), non-ideal MHD simulation ($\alpha=6.8\times10^{-4}$), and pure hydrodynamical simulation (non-turbulence). They measured the accretion rate of particles for an embedded planet with $m=3\times10^{-3}$ and $3\times10^{-2}$ in the Epstein regime. In all cases, they found that pebble accretion occurs efficiently when ${\rm St}=0.1$--1, but the strong turbulence reduces the accretion probability when ${\rm St}\lesssim0.03$ for low mass planet ($m=3\times10^{-3}$). In our study, the reduction of the accretion probability of small pebbles (${\rm St}\lesssim$0.01) can be seen even for $m=0.03$, which corresponds to a high mass planet in \cite{Xu:2017} (\Figref{fig:M-dot-alpha-Ep}a). This may be the result of resolution in the vicinity of the planet. In \cite{Xu:2017}, the Bondi radius is resolved by $\sim1$--3 cells, while it is resolved by at least 40 cells in the radial direction in our study. Since the typical scale of the perturbed region is the Bondi radius \cite[]{Kuwahara:2019}, sufficient resolution may be required to take into account the effects of planet-induced gas flow. However, we emphasize that, in contrast to \cite{Xu:2017} where the turbulent stirring was directly resolved, our model assumed that the turbulent diffusion balances the tide in the vertical direction. In addition, \cite{Xu:2017} assumed $\eta=0.1$ because they focus on the growth of the planet in the outer region of the disk, which may cause the differences in results. We already discussed the effect of the headwind in the previous section.

\cite{Picogna:2018} investigated the accretion of pebbles onto a planet embedded in a 3D globally isothermal disk. They modeled two types of disks: the turbulent disk where the turbulence is driven by hydrodynamic instability (vertical shear instability; VSI) and the laminar viscous disk where the particles experience stochastic kicks. The turbulent parameter corresponds to $\alpha=1.737\times10^{-3}$. They measured the accretion rate of particles in an equilibrium situation in the Epstein regime. The accretion probability of pebbles with ${\rm St}=10^{-4}$--1 reaches $\sim1$--10\% and has a peak at ${\rm St}\sim10^{-2}$ for the planets at 5.2 au with 5 $M_{\oplus}\,(m=0.12)$ and 10 $M_{\oplus}\,(m=0.24)$ both in VSI turbulent and in laminar viscous disks. As shown in Figs. \ref{fig:M-dot-alpha-Ep}b and c, the accretion probabilities have peaks at ${\rm St}\sim0.03$ and fall in the range of $\sim$1--10\% when $\alpha\sim10^{-3}$, which is consistent with \cite{Picogna:2018}. However, the accretion probability for a pebble with ${\rm St}=10^{-3}$ still has the order of $\sim10$\% in the viscous disk in \cite{Picogna:2018}, while it has only $\sim1$\% in our study. In contrast to our study, \cite{Picogna:2018} consider the random motion of the pebbles due to turbulence, which may lead to the high accretion probability of the smaller pebbles.

\subsubsection{Effect of accretion heat}
When the planet accretes solid materials, accretion heating is activated. Given the high accretion luminosity and the baroclinic fluid, spiral-like streamlines rise vertically above the growing planet and form an outflow column escaping the Hill sphere after the gas coming out from between the horseshoe and shear region in the first and third quadrants in the $x$-$y$ plane in the midplane \cite[]{Chrenko:2019}. In contrast to our results, such upward gas flow may suppress the accretion of pebbles coming from high latitudes.

%%%%%%%%%%%%%%%%%%%%%%%%%%%%%%%%%
\subsection{Implication for the formation of planetary systems}
We propose a formation scenario of planetary systems. Firstly we summarize two parameters: the turbulent parameter and the size distribution of the solid materials in a disk, and then give the global structure of the disk in terms of these parameters (Sects. \ref{sec:turbulence} and \ref{sec:size}). Finally we introduce our scenario based on our results (Sect. \ref{sec:scenario}).

\subsubsection{Turbulence in a protoplanetary disk}\label{sec:turbulence}
Pebble accretion depends on the turbulence strength in a disk. One of the possible origins of the disk turbulence is the magneto-rotational instability \cite[MRI,][]{Balbus:1991}. The MRI can develop in a region where the ionization degree of the gas is high and produces strong turbulent intensities of $\alpha\sim10^{-3}$--$10^{-2}$. However, Ohmic diffusion would suppress the MRI, which creates an MRI-inactive region \cite[dead-zone,][]{Gammie:1996}. The dead-zone lies between a few times 0.1 au and about a few tens of 10 au in a disk, in which the $\alpha$ value should be smaller, $\alpha\lesssim10^{-4}$. 

Even if the ionization degree is low, the turbulence is generated through hydrodynamic instabilities: vertical shear instability \cite[VSI,][]{Urpin:1998,Urpin:2003,Nelson:2013,Baker:2015,Lin:2015}, convective overstability \cite[COV,][]{Klahr:2003,Klahr:2014,Lyra:2014,Latter:2016}, subcritical baroclinic instability \cite[SBI,][]{Klahr:2003,Petersen:2007,Lesur:2010}, and zombie vortex instability \cite[ZVI,][]{Marcus:2013,Marcus:2015}. These instabilities develop in different regions of the disk. Assuming a MMSN disk model, the ZVI can develop in the inner region of the disk, $\lesssim1$ au, where the gas is more adiabatic, and produce $\alpha\sim10^{-3}$; the COV can develop in the region 1--10 au and trigger a non-linear state of vortex amplification coined as SBI, which induces $\alpha\sim10^{-4}$--$10^{-2}$; the VSI can develop in the outer region of the disk, $>10$ au, where there tends to be isothermal and generate weak turbulence, $\alpha\sim10^{-4}$ \cite[][and references therein]{Malygin:2017}. Self-regulation may limit the turbulence strength produced by ZVI to be as little as $\alpha\sim10^{-5}$.

The turbulence strength beyond a few tens of au can be constrained observationally. To explain the signature of the millimeter-wave polarization and the ring structure of the HL Tau disk shown by Atacama Large Millimeter/submillimeter Array (ALMA) observations \cite[e.g.,][]{ALMA:2015}, the maximum dust size should be $\sim150$ $\mu$m (${\rm St}\sim10^{-4}$--$10^{-3}$), and the dust grains are settled to the midplane of the disk \cite[]{Kataoka:2016,Okuzumi:2019}. This requires that the $\alpha$ value should be low, $\alpha\lesssim10^{-4}$--$10^{-3}$. Though we adopt the small dust size suggested from the polarization observations of ALMA in the following discussion, we note that it has been argued that a certain proportion of the dust has grown up to $>10$ cm based on the Very Large Array (VLA) image at 1.3 cm. \cite[]{Greaves:2008}. \cite{Dullemond:2018} considered dust trapping in radial pressure bumps as a mechanism to produce dust rings in five different systems observed by the ALMA large program DSHARP campaign. The constrained $\alpha/{\rm St}$ differs in each ring but it is larger than $\sim10^{-2}$. The relatively weak level of turbulence suggested from the observations is also preferable for the formation of ring structures by secular gravitational instability \cite[]{Takahashi:2014,Takahashi:2016,Tominaga:2019}.

The possible origin of the turbulence and the $\alpha$ values vary according to the previous studies. Considering the distribution of the turbulence strength in a disk, we may constrain the types of the planets that are finally formed. The growth of the planets in the strong turbulence region is expected to be slower than that in the weak turbulence region (Figs. \ref{fig:M-dot-alpha} and \ref{fig:M-dot-alpha-Ep}). However, since the accretion probability changes significantly depending on the Stokes number, both $\alpha$ and {\rm St} are needed to discuss planet formation. We discuss the size distribution of the solid materials in a disk in the following section.
%%%%%%%%%%%%%%%%%%%%%%%%%%%%%%%

%%%%%%%%%%%%%%%%%%%%%%%%%%%%%%%%%%%
\subsubsection{Size distribution of pebbles}\label{sec:size}
The planet-induced gas flow significantly reduces the accretion probability of the smaller pebbles (${\rm St}\lesssim10^{-2}$) when pebble accretion occurs in 2D or the planetary mass is small, $m=0.03$ (Figs. \ref{fig:M-dot}--\ref{fig:M-dot-alpha-Ep}). The size distribution of the solid materials in a disk may also regulate the types of the planets that are finally formed.  

\cite{Okuzumi:2019} has shown that the dust particles covered with the \ce{H2O}-ice mantle in the HL Tau disk can grow up to $\sim$50 mm (${\rm St}\sim5\times 10^{-2}$) between the \ce{H2O} and \ce{CO2} snowlines. Inside the \ce{H2O} snowline, where the dust grains exist as non-sticky silicate grains, the maximum dust size is $\sim$1 mm (${\rm St}\sim10^{-3}$). Outside the \ce{CO2} snowline, since the \ce{CO2} ice is as non-sticky as silicate grains \cite[]{Musiolik:2016a,Musiolik:2016b}, the upper limit of the size of the dust is $\sim$0.1--1 mm (${\rm St}\sim10^{-3}$--$10^{-2}$). 

Given such a size distribution, the growth of the planets inside the \ce{H2O} snowline may be expected to be slower than that between the \ce{H2O} and \ce{CO2} snowlines if we consider the reduction of accretion probability of small pebbles due to the planet-induced flow.

\subsubsection{Planet formation via pebble accretion in the planet-induced gas flow}\label{sec:scenario}
Now we introduce the formation scenario of planetary systems based on our results and above discussions (\Figref{fig:summary}). We adopted three hydrodynamic instabilities as the possible origins of the turbulence: ZVI, COV, and VSI. We divided the disk into three sections according to the previous studies and assumed turbulence strength in each section as: $\alpha\sim10^{-5}$ ($\lesssim1$ au), $\alpha\sim10^{-3}$ ($\sim1$--10 au), and $\alpha\sim10^{-4}$ ($\gtrsim10$ au) \cite[]{Malygin:2017,Lyra:2019}. The outermost region ($>100$ au) may be MRI-active, but here we set $\alpha\sim10^{-4}$ from the standpoint of observational constraints. Given the size distribution of the solid materials in a disk \cite[]{Okuzumi:2019}, we assumed that the pebbles have ${\rm St}\sim10^{-3}$ ($\lesssim$1 au), ${\rm St}\sim3\times10^{-2}$ ($\sim$1--10 au), and ${\rm St}\sim3\times10^{-3}$ ($\gtrsim$10 au), which determine the pebble accretion regime. The pebble accretion regime changes according to the pebble size and the distance from the central star (Eqs. (\ref{eq:Epstein-regime}), (\ref{eq:Stokes-regime}), and (\ref{eq:pebble-transition})). Even in the region around $\sim1$ au, pebble accretion occurs in the Epstein regime if ${\rm St}\leq10^{-2}$ (\Equref{eq:pebble-transition}).   

Here we consider the growth of the protoplanet with $m=0.03$, which is the lower limit of planetary mass used in our study. Since we only study the shear regime of pebble accretion, we do not consider the formation of proto-cores or proto-planets via pebble accretion. We propose a possible scenario for the formation of the planetary system as follows:
\begin{enumerate}
\item 
The rocky terrestrial planets or super-Earths are formed in the inner region of the disk, $\lesssim$1 au, where $\alpha$ and pebble size are small. In the region around $\sim1$ au, since the accretion occurs in the Epstein regime (\Equref{eq:pebble-transition}), the accretion probability is $P_{\rm acc}\sim3\times10^{-4}$ when we assumed that $\alpha=10^{-5}$ and ${\rm St}=10^{-3}$ (\Figref{fig:M-dot-alpha-Ep}a). The achieved accretion probability is too small to grow the planets much larger within the typical lifetime of the gas disk ($\sim3$--10 Myr). In this case we would expect that the planets that are finally formed are the rocky terrestrial planets. Within the growth track, they may experience  inward migration or giant impacts. In the region very close to the star ($\lesssim0.1$ au), pebble accretion occurs in the Stokes regime. In this case we would have $P_{\rm acc}\sim10^{-2}$ when we assumed that $\alpha=10^{-5}$ and ${\rm St}=10^{-3}$ (\Figref{fig:M-dot-alpha}a). The accretion regime is switched during inward migration, which may lead to the formation of more massive planets like super-Earths.
\item
The gas giants are formed in the middle region of the disk, $\sim1$--10 au, where $\alpha$ and pebble size becomes larger than those in the inner region. In this region, since the accretion occurs in the Epstein regime (\Equref{eq:pebble-transition}), the accretion probability is $P_{\rm acc}\sim3\times10^{-2}$ when we assumed that $\alpha=10^{-3}$ and ${\rm St}=3\times10^{-2}$ (\Figref{fig:M-dot-alpha-Ep}a). As the planets grow, the accretion probability becomes larger. We would expect that the planets might exceed the critical core mass ($>10\ M_{\oplus}$) within the typical lifetime of the disk.
\item
The ice giants are formed in the outer region. Since the accretion occurs in the Epstein regime (\Equref{eq:pebble-transition}), the accretion probability is $P_{\rm acc}\sim9\times10^{-3}$ when we assumed that $\alpha=10^{-4}$ and ${\rm St}=3\times10^{-3}$ (\Figref{fig:M-dot-alpha-Ep}a). With the accretion of a moderate amount of icy pebbles, the planet would eventually become ice giants without evolving into gas giants. 
\end{enumerate}
Our scenario may be a natural explanation of the occurrence rate of the exoplanets: an abundance of super-Earths at $<1$ au \cite[]{Fressin:2013,Weiss:2014} and a possible peak in the occurrence of gas giants at $\sim2$--3 au \cite[]{Johnson:2010,Fernandes:2019}, as well as the architecture of the solar system. 

The dichotomy between the inner super-Earths and outer gas giants may be the result of the reduction of the pebble isolation mass (\Equref{eq:peb-iso}) in the inner region of the inviscid disk \cite[]{Fung:2018}. However, the small pebbles (${\rm St}\lesssim10^{-3}$) may not be trapped at the local pressure maxima, and they continue to contribute to the growth of the planet \cite[]{Bitsch:2018}. Since the accretion probability of small pebbles decreases significantly in the inner region of the disk, our scenario has the potential to explain the dichotomy even if the influx of small pebbles does not stop at the pressure maxima.

The spacial variety of turbulence strength and pebble size may induce the formation of planetary seeds. Grains would accumulate where the turbulence strength or pebble size changes and the pile-up may trigger planetesimal formation via direct growth and streaming instability \cite[]{Kretke:2007,Youdin:2005,Youdin-Johansen:2007,Johansen:2007,Morbidelli:2015,Ida:2016}.

%---------------------------------------------------------
%---------------------------------------------------------
%---------------------------------------------------------

\section{Conclusions} \label{sec:conclusion}
We investigated the influence of the 3D planet-induced gas flow on pebble accretion. We considered non-isothermal, inviscid gas flow,
and performed a series of 3D hydrodynamical simulations on
a spherical polar grid that has a planet placed at its center. Then we numerically integrated the equation of the motion of pebbles in 3D  using hydrodynamical simulation data, which included the Coriolis, tidal, two-body interaction, and gas drag forces. Three-types of orbital calculation of pebbles are conducted in this study: in the unperturbed shear flow (Shear case), in the planet-induced gas flow in the Stokes regime (PI-Stokes case), and in the planet-induced gas flow in the Epstein regime (PI-Epstein case). The subject of the range of dimensionless planetary mass in this study is $m=0.03$--0.3, which corresponds to a size ranging from three Mars masses to a super-Earth-sized planet, $M_{\rm pl}=0.36$--3.6 $M_{\oplus}$, orbiting a solar-mass star at 1 au. Since the planetary masses are larger than the transition mass, we only considered the shear regime of pebble accretion. We summarize our main findings as follows.
\begin{enumerate}
\item
The trajectories of pebbles in the planet-induced gas flow differ significantly from those in the unperturbed shear flow, in particular for the pebbles with ${\rm St}\leq10^{-1}$ when $m=0.1$ (\Figref{fig:m01-peb-line}). Most pebbles coming from within the vicinity of the planetary orbit move away from the planet along the horseshoe flows, which is consistent with a previous study \cite[]{Popovas:2018a}. The outflow of the gas deflects the pebble trajectories and inhibits small pebbles from accreting. The 3D trajectories of pebbles also show a similar trend as seen in 2D because the horseshoe streamlines extend in the vertical direction, keeping its configurations, and the vertical scale of the outflow is $\sim R_{\rm Bondi}$ (\Figref{fig:peb-line-3D}). The region that is perturbed by the gravity of the planet can be scaled by the Bondi radius. Therefore, even if the pebbles with a certain Stokes number are hardly influenced by the planet-induced gas flow when the planetary mass is small, the same-sized pebbles are highly influenced when the planetary mass is large (\Figref{fig:m003-03-peb-line}).
\item
The width of the accretion window (\Equref{eq:acc-width}) and the accretion cross section (\Equref{eq:acc-sec}) in the planet-induced gas flow becomes smaller than those in the unperturbed shear flow (Figs. \ref{fig:impact-parameter} and \ref{fig:integrate-cross-section}). The reduction of these quantities becomes more prominent when the Stokes number is small, or when in the Epstein regime rather than the Stokes regime.
\item
The accretion probability of pebbles (\Equref{eq:Pacc}) in the PI-Stokes case matches with that in the Shear case when ${\rm St}\gtrsim3\times10^{-3}$--$10^{-2}$, or decreases when ${\rm St}$ falls below the preceding value in 2D. Whereas the accretion probability matches or is slightly larger than that in the Shear case when $m\geq0.1$, or smaller when $m=0.03$ for ${\rm St}\lesssim3\times10^{-2}$ in 3D. In the planet-induced gas flow, the accretion cross section shifts to the right as a whole (\Figref{fig:cross-section}), which leads to the increase of the relative velocity between the pebbles and the planet. The reduction of the accretion cross section and the increase of relative velocity cancel each other out. Thus the suppression of pebble accretion in the PI-Stokes case can be seen only when the Stokes number is small, ${\rm St}\sim10^{-3}$, in 2D or the planetary mass is small, $m=0.03$, in 3D. In the Epstein regime, however, since the accretion cross section becomes more significant than those in the PI-Stokes case, the increase of the relative velocity does not fully offset its reduction. Therefore, the accretion probability in the PI-Epstein case becomes smaller than that in the Shear case both in the 2D and in the 3D regardless of assumed ${\rm St}$ and $m$. 
\end{enumerate}
Assuming the global structure of the disk in terms of the distribution of the turbulent parameter and the size distribution of the solid materials in a disk based on previous studies \cite[]{Malygin:2017,Lyra:2019,Okuzumi:2019}, we proposed a formation scenario of planetary systems (\Figref{fig:summary}). The 3D planet-induced gas flow does affect the pebble accretion and may be helpful to explain the distribution of exoplanets (the dominance of super-Earths at $<1$ au \cite[]{Fressin:2013,Weiss:2014}) and a possible peak in the occurrence of gas giants at $\sim2$--3 au \cite[]{Johnson:2010,Fernandes:2019}, as well as the architecture of the solar system. 
%-------------------------------------------------------------------------------------------
\begin{acknowledgements}
We would like to thank an anonymous referee for constructive comments. We thank Athena++ developers: James M. Stone, Kengo Tomida, and Christopher White. This study has greatly benefited from fruitful discussions with Satoshi Okuzumi. Numerical computations were in part carried out on Cray XC30 at the Earth-Life Science Institute and on Cray XC50 at the Center for Computational Astrophysics at the National Astronomical Observatory of Japan.  This work was supported by JSPS KAKENHI Grant number 17H01175, 17H06457, 18K13602, 19H01960, 19H05072. 
\end{acknowledgements}

%-------------------------------------------------------------------------------------------

%% references
%%\raggedright              %% only for adsaa with dvips, not for pdflatex

\appendix

\label{sec:appendix}
\section{Analytical estimation of pebble accretion}\label{sec:pebbleaccretion}
\subsection{Width of accretion window and accretion cross section}
The width of the accretion window in the unperturbed shear flow is expressed by
\begin{align}
b_{x}=b_{x,0}\exp\left[-\left(\frac{{\rm St}}{2}\right)^{0.65}\right],\label{eq:bx}
\end{align}
where the exponential factor is the cutoff parameter given by \cite{Ormel:2012} and $b_{x,0}$ is the solution to \cite[]{Ormel:2010}
\begin{align}
b^{3}_{x,0}+\frac{2}{3}v_{\rm hw}b^{2}_{x,0}-\frac{8}{3}m{\rm St}=0,
\end{align} 
which can be divided into two formulas as follows \cite[]{Ormel:2010,Lambrechts:2012,Guillot:2014,Ida:2016,Sato:2016}:
\begin{empheq}[left={b_{x,0}\simeq\empheqlbrace}]{alignat=2}
&\displaystyle2\sqrt{\frac{{\rm St}}{\tau_{\rm B}}}\ R_{\rm Bondi}, \quad &(M_{\rm pl}< M_{\rm t}:  \text{headwind regime})\label{eq:b-hw}\\
&\displaystyle2{\rm St}^{1/3}\ R_{\rm Hill},  \quad &(M_{\rm pl}> M_{\rm t}:\text{shear regime})\label{eq:b-sh}
\end{empheq}
where $\tau_{\rm B}$ is the gas-free core-crossing time associated with the Bondi radius,
\begin{align}
\tau_{\rm B}=\frac{R_{\rm Bondi}}{v_{\rm hw}}.
\end{align}
Again, we only considered the shear regime in this study, since the masses of the embedded planets (\Equref{eq:planetarymass}) are much larger than the transition mass (\Equref{eq:transition}).

We derived the height of the accretion window with a similar procedure of the derivation of $b_{x,0}$ in \cite{Ormel:2010}. The gravitational force acting on the pebble at $z=b_{z,o}$ is $\bm{F}_{\rm grav}=m/b^{2}_{z,0}$. In the shear regime, the approach velocity of pebbles can be written by $\bm{v}_{\rm p,\infty}=-3/2x\bm{e}_{y}$, which gives the interaction timescale as $t_{\rm int}=b_{z,0}/v_{\rm p,\infty}$. Gas drag can be neglected if $t_{\rm int}<{\rm St}$, but otherwise cannot. When $t_{\rm int}>{\rm St}$ the pebble velocity equilibrates to $\bm{v}_{\rm p}=\bm{F}_{\rm drag}{\rm St}$. According to \cite{Ormel:2010}, we equate $\bm{v}_{\rm p}$ with $\bm{v}_{\rm p,\infty}/4$, and then obtain
\begin{align}
b_{z,0}=\sqrt{\frac{8m{\rm St}}{3x}}.
\end{align}
Applying the same cutoff parameter in \Equref{eq:bx}, the height of the accretion window is described by
\begin{align}
b_{z}=b_{z,0}\exp\left[-\left(\frac{{\rm St}}{2}\right)^{0.65}\right].\label{eq:bz}
\end{align}
From Eqs. (\ref{eq:bx}) and (\ref{eq:bz}), we calculate the accretion cross section by 
\begin{align}
A_{\rm acc}=2\int_{0}^{b_{z}}\int_{0}^{b_{x}}dxdz=16\,{\rm St}^{2/3}R_{\rm Hill}^{2}\exp\left[-\frac{3}{2}\left(\frac{{\rm St}}{2}\right)^{0.65}\right].\label{eq:acc}
\end{align}
Equations \ref{eq:bx} and \ref{eq:acc} are plotted in Figs. \ref{fig:impact-parameter} and \ref{fig:integrate-cross-section} with dashed-dotted lines. Both equations show excellent agreement with the numerically calculated width of the accretion window and accretion cross section in the Shear case.
%To accrete the pebble, two conditions are required: (1) the stopping time of the pebble must be smaller than the duration of the time that the pebble experience the gravity of the planet , $t_{\rm enc}=2b/v_{{\rm p},\infty}$ (\textit{encounter time}), where $v_{{\rm p},\infty}$ is the unperturbed approach velocity of the pebble. (2) the duration of the time that the pebble to settle to the planet, $t_{\rm settl}=b/v_{\rm settle}=b^{3}/m{\rm St}$ (\textit{settling time}), where $v_{\rm settle}$ is the terminal velocity uof pebble relative to the planet, must be smaller than the encounter time. %The first condition is always satisfied when we consider ${\rm St}\leq1$ because $v_{{\rm p},\infty}\sim3b/2$ in the shear regime.
\subsection{Pebble isolation mass}
When a growing planet opens up a gap in a disk and generates a pressure maxima outside of its orbit, pebble accretion halts \cite[]{Lambrechts:2014}. This happens when the core mass reaches pebble isolation mass: 
\begin{align}
M^{L14}_{\rm iso}\approx20\left(\frac{a}{5\ {\rm au}}\right)^{3/4}\ M_{\oplus}.
\end{align}
A subsequent study derived a detailed pebble isolation mass,
\begin{align}
M_{\rm iso}=25\ f_{\rm fit} M_{\oplus}+\frac{\Pi_{\rm crit}}{\lambda}\ M_{\oplus},\label{eq:peb-iso}
\end{align}
where $\lambda\approx0.00476/f_{\rm fit}$, $\Pi_{\rm crit}=\alpha/2{\rm St}$, and
\begin{align}
f_{\rm fit}=\Biggl[\frac{H/a}{0.05}\Biggr]^{3}\Biggl[0.34\ \Biggl(\frac{3}{\log(\alpha)}\Biggr)^{4}+0.66\Biggr]\Biggl[1-\frac{\frac{\partial \ln P}{\partial \ln a}+2.5}{6}\Biggr],
\end{align}
\cite[]{Bitsch:2018}.

\subsection{Accretion rate}
The accretion rate of pebbles, $\dot{M}_{\rm p}$, is given analytically by \cite[]{Ormel-note}:
\begin{empheq}[left={\dot{M}_{\rm p}=\empheqlbrace}]{alignat=2}
&\dot{M}_{\rm 2D}, \quad &(\text{2D, analytical})\label{eq:M-dot-2D}\\
&\displaystyle\dot{M}_{\rm 2D}\frac{2b_{x}}{b_{x}+H_{\rm p}\sqrt{8/\pi}}, \quad &(\text{3D, analytical}) \label{eq:M-dot-3D}
\end{empheq}
where $\dot{M}_{\rm 2D}$ is calculated by
\begin{align}
\dot{M}_{\rm 2D}=2\int_{0}^{b_{x}}\Sigma_{\rm p}\bm{v}_{\rm p,\infty}dx.
\end{align}
We use a slightly different formula of 3D accretion rate described in \cite{Ormel:2017}, where the factor of 2 is not included. Equations \ref{eq:M-dot-2D} and \ref{eq:M-dot-3D} are plotted in \Figref{fig:M-dot} with dashed-dotted lines. Both equations show excellent agreement with the numerically calculated accretion rate in the Shear case.

%%%%%%%%%%%%%%%%%%%%%%%%%%%%%%
%%%%%%%%%%%%%%%%%%%%%%%%%%%%%% 
\section{Interpolation of simulation data}\label{sec:interpolation}
In a series of numerical results obtained from our grid simulations using Athena++ code, all physical quantities were given as discrete data on the center of each grid. In this study, we interpolated the gas velocity and the density at the position using the bilinear interpolation method. We note that when we interpolate the gas velocity and the density, since the hydro simulations are performed in a spherical polar coordinates, we transformed the position of the pebble in Cartesian coordinates to that in spherical coordinates. Given the position of the pebble at an arbitrary time, $\bm{r}=(x,y,z)=\left(r(t),\theta(t),\phi(t)\right)$, a certain physical quantity at the position of the pebble, $q(\bm{r},t)$,  is described as the following interpolation formula:
\begin{align}
q(\bm{r},t)=(1-\zeta)q_{{\rm itp},1}+\zeta q_{{\rm itp},2},
\end{align}
where $q_{{\rm itp},1}$ and $q_{{\rm itp},2}$ are the interpolated physical quantities on the top and bottom surface of the cell including the pebble,
\begin{align}
q_{{\rm itp},1}=\left(1-\eta\right)\left(1-\xi\right)&q_{i,j,k}+\eta\left(1-\xi\right)q_{i+1,j, k}\\ \nonumber
&+\eta\xi q_{i+1,j, k+1}+\left(1-\eta\right)\xi q_{i,j.k+1}, \\
q_{{\rm itp},2}=\left(1-\eta\right)\left(1-\xi\right)&q_{i,j+1,k}+\eta\left(1-\xi\right)q_{i+1,j+1, k}\\ \nonumber
&+\eta\xi q_{i+1,j+1, k+1}+\left(1-\eta\right)\xi q_{i,j+1,k+1},
\end{align}
where $q_{i,j,k}$ is a certain physical quantity at the center of the grid, the subscripts denote the grid number, and $\zeta,\eta,\xi$ are given by
\begin{align}
\zeta = \frac{\theta(t)-\theta_{j}}{\theta_{j+1}-\theta_{j}},\ \ 
\eta =\frac{r(t)-r_{i}}{r_{i+1}-r_{i}},\ \ \xi = \frac{\phi(t)-\phi_{k}}{\phi_{k+1}-\phi_{k}}.
\end{align}
After the interpolation, we integrate \Equref{eq:EOM} in Cartesian coordinates. Then we have the gas velocity after being converted from spherical polar coordinates to Cartesian coordinates using Athena ++.

\section{Independence of accretion rate from orbital radius}\label{sec:appendix-c}
From Eqs. (\ref{eq:vhw}), (\ref{eq:eta}), (\ref{eq:M-dot-disk}), and (\ref{eq:vdrift}), the surface density of pebbles can be expressed by 
\begin{align}
\Sigma_{\rm p}=\frac{\dot{M}_{\rm disk}}{2\pi\times\frac{\mathrm{d}\ln P}{\mathrm{d}\ln a}}\frac{1+{\rm St}^{2}}{\rm St}\frac{1}{H^{2}\Omega}.\label{eq:c1}
\end{align}
When $P\propto a^{-q}$, where $q$ is an arbitrary index, which holds for typical disk models including the MMSN model, the pressure gradient, $\mathrm{d}\ln P/\mathrm{d}\ln a$, on the right-hand side of \Equref{eq:c1} becomes constant. Therefore, 
\begin{align}
\Sigma_{\rm p}\propto H^{-2}\Omega^{-1}.\label{eq:c2}
\end{align}
From Eqs. (\ref{eq:pebble-density}), (\ref{eq:pebble-scaleheight}), and (\ref{eq:c2}), the pebble density can be described by
\begin{align}
\rho_{\rm p}\propto \frac{\Sigma_{\rm p}}{H}\propto H^{-3}\Omega^{-1}.\label{eq:c3}
\end{align}
In this study, the length scale is normalized by the disk scale height, then the integrals in Eqs. (\ref{eq:M-dot-2D-sim}) and (\ref{eq:M-dot-sim}) are proportional to $H$ and $H^{2}$, respectively, when the dimensionless planetary mass, $m$, is fixed. Thus, from Eqs. (\ref{eq:M-dot-2D-sim}), (\ref{eq:M-dot-sim}), (\ref{eq:c2}), and (\ref{eq:c3}), the accretion rates in both 2D and 3D are proportional to 
\begin{align}
&\dot{M}_{\rm p,\,2D}\propto\Sigma_{\rm p}v_{\rm p,\infty}H\propto H^{-2}\Omega^{-1}\times H^{2}\Omega=const.,\label{eq:c4}\\
&\dot{M}_{\rm p,\,3D}\propto\rho_{\rm p}v_{\rm p,\infty}H^{2}\propto H^{-3}\Omega^{-1}\times H^{3}\Omega=const.\label{eq:c5}
\end{align}
The accretion rates both in 2D and 3D are independent of the orbital radius, $a$, for a fixed dimensionless planetary mass, $m$. 

\end{document}